\documentclass[fleqn,usenatbib]{mnras}

\usepackage{newtxtext,newtxmath}
\usepackage{natbib}
\usepackage[T1]{fontenc}

\DeclareRobustCommand{\VAN}[3]{#2}
\let\VANthebibliography\thebibliography
\def\thebibliography{\DeclareRobustCommand{\VAN}[3]{##3}\VANthebibliography}

\usepackage{graphicx}	% Including figure files
\usepackage{amsmath}	% Advanced maths commands
\usepackage{booktabs}

\newcommand{\UV}{\mathrm{UV_{275}}}
\newcommand{\U}{\mathrm{U_{336}}}
\newcommand{\B}{\mathrm{B_{438}}}

\newcommand{\I}{\mathrm{I_{814}}}
\newcommand{\bp}{\mathrm{BP}}
\newcommand{\rp}{\mathrm{RP}}

\newcommand{\BIB}{(\B-\I,\B)}
\newcommand{\BII}{(\B-\I,\I)}
\newcommand{\UBB}{(\U-\B,\B)}
\newcommand{\UVUU}{(\UV-\U,\U)}

\newcommand{\lrb}[1]{\left(#1\right)}

\newcommand{\Msun}{\mathrm{M_{\sun}}}

\newcommand{\msunyr}{\Msun~\mathrm{yr^{-1}}}
\newcommand{\gaia}{{\it Gaia}}
\newcommand{\hst}{{\it HST}}
\newcommand{\chandra}{{\it Chandra}}
\newcommand{\vla}{{\it VLA}}

\newcommand{\kpc}{\mathrm{kpc}\,}

\newcommand{\msun}{\mathrm{M_\odot}}

\newcommand{\nodata}{$\cdots$}
\newcommand{\nh}{N_\mathrm{H}}

\newcommand{\rh}{r_\mathrm{h}}
\newcommand{\rc}{r_\mathrm{c}}
\newcommand{\mas}{\mathrm{mas}}
\newcommand{\ks}{\mathrm{ks}}
\newcommand{\kev}{\mathrm{keV}}
\newcommand{\wavdetect}{{\tt wavdetect}}
\newcommand{\ghz}{\mathrm{GHz}}
\newcommand{\nulo}{\nu_\mathrm{low}}
\newcommand{\nuhi}{\nu_\mathrm{high}}
\newcommand{\slow}{S_\mathrm{low}}
\newcommand{\shigh}{S_\mathrm{high}}
\newcommand{\mujy}{\mu\mathrm{Jy}}
\newcommand{\mujyperbeam}{\mu\mathrm{Jy~beam^{-1}}}
\newcommand{\doff}{D_\mathrm{off}}
\newcommand{\perr}{r_\mathrm{err, x}}
\newcommand{\perrradio}{r_\mathrm{err, r}}
\newcommand{\nc}{N_c}

\newcommand{\ncounterpartstox}{6}
\newcommand{\nradiosource}{14}
\newcommand{\probagn}{P_\mathrm{AGN}}

\newcommand{\wstat}{{\tt wstat}}
\newcommand{\pow}{{\tt pl}}
\newcommand{\bbody}{{\tt bbody}}
\newcommand{\apec}{{\tt apec}}
\newcommand{\fsoft}{f_\mathrm{0.5-2}}
\newcommand{\fhard}{f_\mathrm{2-7}}
\newcommand{\ergscm}{\mathrm{erg~s^{-1}~cm^{-2}}}
\newcommand{\fbroad}{f_\mathrm{1-10}}
\newcommand{\rbb}{R_\mathrm{bb}}
\newcommand{\ergps}{\mathrm{erg~s^{-1}}}
\newcommand{\mv}{M_\mathrm{V}}
\newcommand{\mb}{M_\mathrm{B}}
\newcommand{\mi}{M_\mathrm{I}}
\newcommand{\revise}[1]{{#1}}
\newcommand{\referee}[1]{{\color{black}{#1}}}

\defcitealias{Bahramian20}{B20}
\defcitealias{Shishkovsky20}{Sh20}
\defcitealias{DAntona22}{D22}

\title[Exotic binaries in M14]{Exploration of faint X-ray and radio sources in the massive globular cluster M14: A UV-bright counterpart to Nova Ophiuchus 1938}

\author[Yue Zhao et al.]{Yue Zhao,$^{1}$\thanks{E-mail: yue.zhao@soton.ac.uk}
Francesca D'Antona,$^{2}$
Antonino P. Milone,$^{3,4}$
Craig Heinke,$^{5}$
Jiaqi Zhao,$^{5}$
\newauthor
Phyllis Lugger,$^{6}$
and Haldan Cohn$^{6}$
\newauthor
\\
$^{1}$School of Physics \& Astronomy, University of Southampton, Highfield, Southampton SO17 1BJ, UK\\
$^{2}$INAF -- Osservatorio Astronomico di Roma, Via Frascati 33, I-00040 Monte Porzio Catone, Roma, Italy\\
$^{3}$Dipartimento di Fisica e Astronomia ‘Galileo Galilei’, Univ. di Padova, Vicolo dell’Osservatorio 3, I-35122 Padova, Italy\\
$^{4}$Istituto Nazionale di Astrofisica – Osservatorio Astronomico di Padova, Vicolo dell’Osservatorio 5, I-35122 Padova, Italy\\
$^{5}$Department of Physics, University of Alberta, Edmonton, AB T6G 2E1, Canada\\
$^{6}$Department of Astronomy, Indiana University, 727 E. Third St.,
Bloomington, IN 47405, USA\\
}

\date{Accepted XXX. Received YYY; in original form ZZZ}

\pubyear{2024}

\begin{document}

\label{firstpage}
\pagerange{\pageref{firstpage}--\pageref{lastpage}}
\maketitle

\begin{abstract}
Using a 12\, ks archival \chandra\ X-ray Observatory ACIS-S observation on the massive globular cluster (GC) M14, we detect a total of $7$ faint X-ray sources within its half-light radius at a $0.5-7\,\kev$ depth of $2.5\times 10^{31}\,\ergps$. We cross-match the X-ray source positions with a catalogue of the {\it Very Large Array} radio point sources and a {\it Hubble Space Telescope} (\hst) UV/optical/near-IR photometry catalogue, revealing radio counterparts to 2 and \hst\ counterparts to 6 of the X-ray sources. In addition, we also identify a radio source with the recently discovered millisecond pulsar PSR 1737$-$0314A. The brightest X-ray source, CX1, appears to be consistent with the nominal position of the classic nova Ophiuchi 1938 (Oph 1938), and both Oph 1938 and CX1 are consistent with a UV-bright variable \hst\ counterpart, which we argue to be the source of the nova eruption in 1938. This makes Oph 1938 the second classic nova recovered in a Galactic GC since Nova T Scorpii in M80. CX2 is consistent with the steep-spectrum radio source VLA8, which unambiguously matches a faint blue source; the steepness of VLA8 is suggestive of a pulsar nature, possibly a \referee{transitional millisecond pulsar} with a late K dwarf companion, though an active galactic nucleus (AGN) cannot be ruled out. The other counterparts to the X-ray sources are all suggestive of chromospherically active binaries or background AGNs, so their nature requires further membership information. 
\end{abstract}


%%%%%%%%%%%%%%%%%%%%%%%%%%%%%%%%%%%%%%%%%%%%%%%%%%
\section{Introduction}
Globular clusters (GCs) have been recognised as veritable factories of close binaries. Throughout their advanced ages, they had witnessed the deaths of the most massive stars that leave behind populations of white dwarfs (WDs), neutron stars (NSs), and black holes (BHs). These compact objects join many dynamical encounters that are facilitated by the very dense GC environment, giving rise to a variety of close binaries hosting compact objects \citep[e.g.,][]{Clark75b, Fabian75, Sutantyo75, Hills76, Camilo05, Ivanova06, Ivanova08}. 

These close binaries could be sources of X-rays. The early X-ray missions (e.g., {\it Uhuru}, {\it OSO-7}) revealed that GCs are orders of magnitude more abundant in X-ray sources compared to the field \citep{Katz75}. Owing to the limited instrument sensitivity, these sources are typically bright \citep[$\geq 10^{36}\,\ergps$;][]{Giacconi74} and are attributed to accreting NSs in low-mass X-ray binaries \citep[LMXBs; see e.g.,][]{Clark75a, Canizares75, Katz75}. Subsequently, more fainter ($\leq 10^{34}\,\ergps$) sources were revealed by the {\it Einstein Observatory} \citep{Hertz83} and {\it ROSAT} \citep{Verbunt01}, and finally till today, the {\it Chandra X-ray Observatory} still provides the unprecendented angular resolution and sensitivity to further push the depths of the observations, revealing a plethora of faint sources in many GCs \citep[e.g.,][]{Grindlay01, Pooley02b, Bassa04,  
%Gendre03, -XMM not Chandra
Heinke05, Kong06, Bassa08, Servillat08, Haggard09, Henleywillis18, Cohn21, Lugger17, Lugger23, Zhao19, Zhao20b,Vurgun22}.

Faint sources have been long suggested to be a mix of multiple types of close binaries. Typically, many GCs host a population of cataclysmic variables (CVs) where a white dwarf accretes from a low-mass donor star \citep[e.g.,][]{Hertz83, Pooley02a, RiveraSandoval18}, while NSs in quiescent LMXBs \citep[qLMXB; e.g.,][]{Verbunt84, Heinke14} and radio millisecond pulsars \citep[MSPs; e.g.,][]{Saito97, Bogdanov10} have also been noted to emit X-rays through thermal and non-thermal processes. %Moreover, it has also been suggested that 
GCs also host a significant population of close binaries made of non-degenerate companions \citep{Bailyn90,Grindlay01}. These close binaries have tidally-locked orbits that force fast stellar rotation, and as a result, they have significantly enhanced chromospheric activity that emits X-rays at an observable level \citep{Dempsey97}, hereafter referred to as active binaries (ABs). To minimise the effect of crowding in GCs, identification of these faint X-ray sources have been greatly aided by incorporating deep and high-resolution imaging observation in UV/optical/near-IR \citep[e.g.,][]{Bassa04, Dieball10, Lugger17, Cohn21, Zhao19, Zhao20b}, and/or in radio \citep[e.g.,][]{Fruchter00,Strader12, Chomiuk13, Shishkovksy18, Zhao20a, Lugger23}.

M14 is one of the most massive GCs in our Galaxy (8th most massive in \citealt{Baumgardt18}, 13th most massive in Baumgardt's 4th cluster update\footnote{\url{https://people.smp.uq.edu.au/HolgerBaumgardt/globular/}},  at $6\times10^5\,\Msun$). It has a distance slightly above average for GCs \citep[$9.3\,\kpc$;][2010 edition]{Harris96}. 
Its relatively low central density (log $\rho_c=3.32$, \citealt{Baumgardt18}) and large core (2.28 pc) combine to give it a relatively low stellar encounter rate (\referee{about a factor of 8 below that of 47 Tuc}; \citealt{Bahramian13}), which suggests it should have only a moderate number of compact binaries, such as quiescent neutron star LMXBs and MSPs, produced by dynamical interactions (e.g. \citealt{Pooley03,Bahramian13}). Indeed, recently five MSPs, a moderate number, have been discovered in M14 using the {\it Five-hundred-meter Aperture Spherical radio Telescope} \citep[FAST;][]{Pan21}; one of these MSPs has a timing solution and a faint X-ray counterpart \citep{ZhaoJ22}. However, the number of close binaries containing BHs is not expected to scale with stellar encounter rate, as the BH population self-segregates early in the cluster history, and strongly influences the cluster parameters by heating the core \citep[e.g.,][]{Kremer18, ArcaSedda18}. \referee{With N-body simulations,} \cite{Ye19} show that MSPs are favoured by larger GC masses but are suppressed by retained BHs; typically, MSPs are more concentrated in GCs that have a smaller number of BHs. M14, as a massive cluster with a large core, seems likely to have numerous BH binaries.

This makes the identification of a number of excess radio sources in M14 of particular interest. \citet{Shishkovsky20} found that M14 has an excess of radio sources within its half-mass radius of 7.3$\pm$3.8 sources. \citet{Zhao21} used a radio luminosity limit of $5\times10^{27}$ erg s$^{-1}$ to compare clusters, and found M14 to have by far the largest radio source excess within its half-mass radius, 14 sources when only 5.7 are expected (nearly a 3$\sigma$ excess, by \citealt{Gehrels86}). The other two clusters that \citet{Zhao21} found to have radio source excesses, M62 and NGC 6440, have high stellar encounter rates. M14, in contrast, may be a window on massive clusters with large numbers of BHs. 

This work %is based on an --Original wording sounded like Chandra was central and other data only supporting; rewriting to put all 3 datasets closer to equal importance.
uses an 
archival \chandra\ X-ray observation of M14, %incorporating 
UV/optical/near-IR \hst\ imaging, and deep \vla\ radio imaging observations %for counterpart identification. 
to identify exotic binaries. 
The paper is organised as follow: Sec \ref{sec:observations} describes observational data used in this study; in Sec \ref{sec:methods}, we show related methods for processing and analysis; in Sec \ref{sec:results}, we present results and discussions on individual sources; and finally in Sec \ref{sec:conclusions}, we draw conclusions.

\section{Observations}
\label{sec:observations}
\subsection{X-ray Observation}
\label{sec:x-ray-observations}
 M14 was observed in 2008-05-24 by the {\it Chandra X-ray Observatory} (Cycle 09; observation ID: 8947; PI Pooley). A single exposure of $12.09\,\ks$ was performed using the ACIS-S detector. We retrieved data of this exposure using the \chandra\ search portal\footnote{\url{http://cda.harvard.edu/chaser/}} and reprocessed the observation files with the up-to-date calibration database (CALDB, version 4.10.4) that is integrated in the {\tt chandra\_repro} script in the {\sc Chandra Interactive Analysis of Observations} ({\sc ciao}) software (version 4.15.1)\footnote{\url{https://cxc.cfa.harvard.edu/ciao/}}.

\begin{figure*}
    \centering
    \includegraphics[scale=0.4]{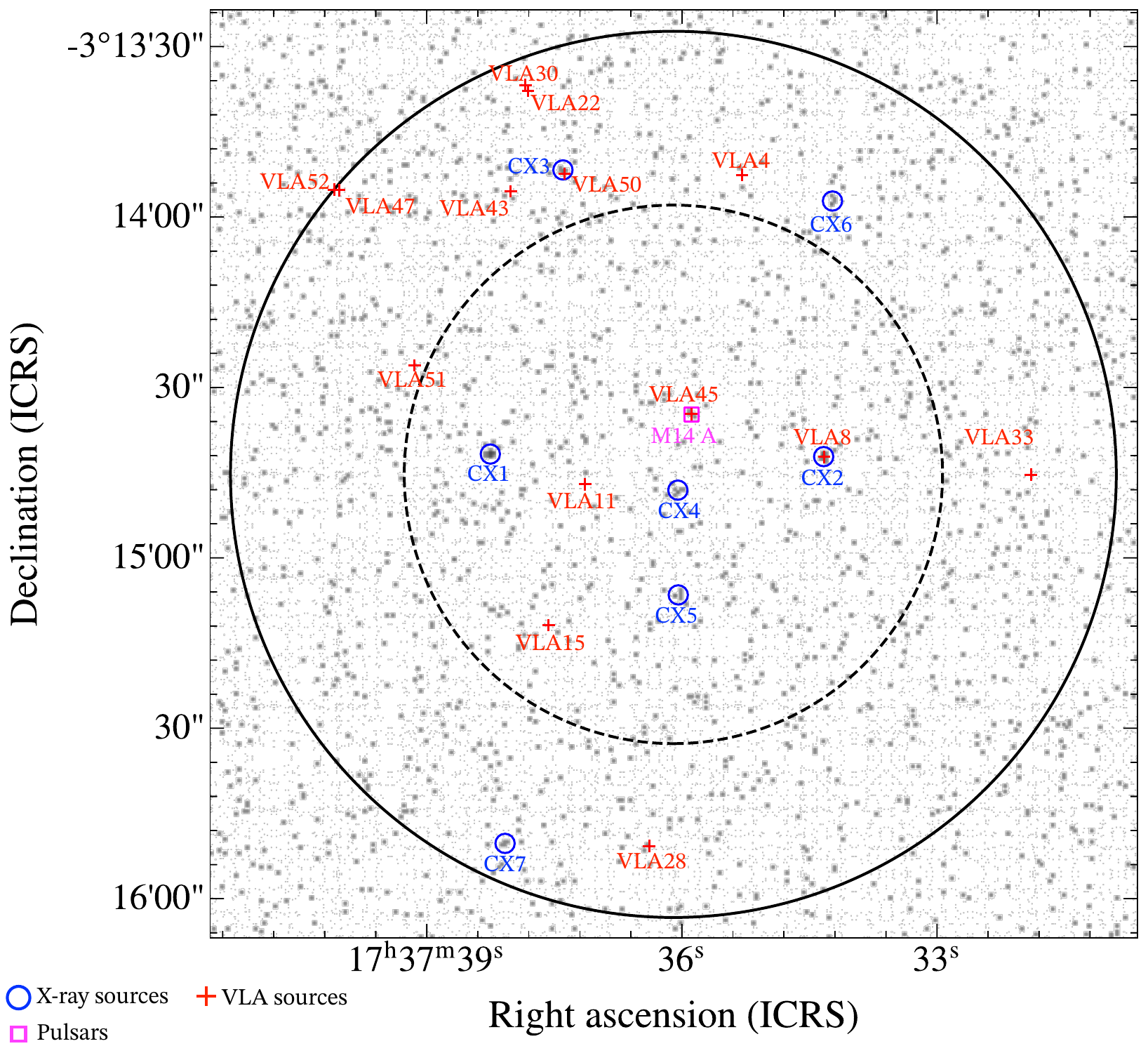}
    \caption{$2\farcm72 \times 2\farcm72$ ACIS-S image of M14 between $0.5$--$10\,\kev$; north is up and east is to the left. Solid and dashed circles depict the $1\farcm3$ half-light radius and the $0\farcs79$ core radius, respectively. X-ray sources found by \wavdetect\ are indicated by blue circles; VLA source positions are indicated by red crosses; and the radio timing position of the known MSP, M14 A, is marked by a magenta square.}
    \label{fig:x_ray_image}
\end{figure*}

\subsection{Radio Observations}
\label{sec:radio-observations}
M14 was observed by {\it the Karl G. Jansky Very Large Array} (\vla) in July 2015 (Project code: 15A-100; PI Strader) as a part of the Milky Way ATCA\footnote{\it The Australia Telescope Compact Array} and VLA Exploration of Radio sources In Clusters (MAVERIC) survey \citep{Tremou18, Shishkovsky20, Tudor22}, a deep radio imaging survey dedicated to finding more accreting compact objects in GCs. The observations of M14 were  performed in the most extended A configuration, totalling 10 hours (8.4 hours on source) of integration. Data were acquired with the C band receiver, which is further split into two 2-GHz sub-bands centered at 5.3 and 7.2$\,\ghz$. For convenience, we use ``low" and ``high" to refer to the 5.3 and 7.2 sub-band, so $\nulo$ and $\nuhi$ denote their central frequencies, and $\slow$ and $\shigh$ represent specific flux densities at $\nulo$ and $\nuhi$, respectively. Fluxes of the two sub-bands can be used to compute the radio spectral index $\alpha$, defined as $S_\nu \propto \nu^\alpha$, where $S_\nu$ is the specific flux density at frequency $\nu$.

\begin{figure}
    \centering
    \includegraphics[width=\columnwidth]{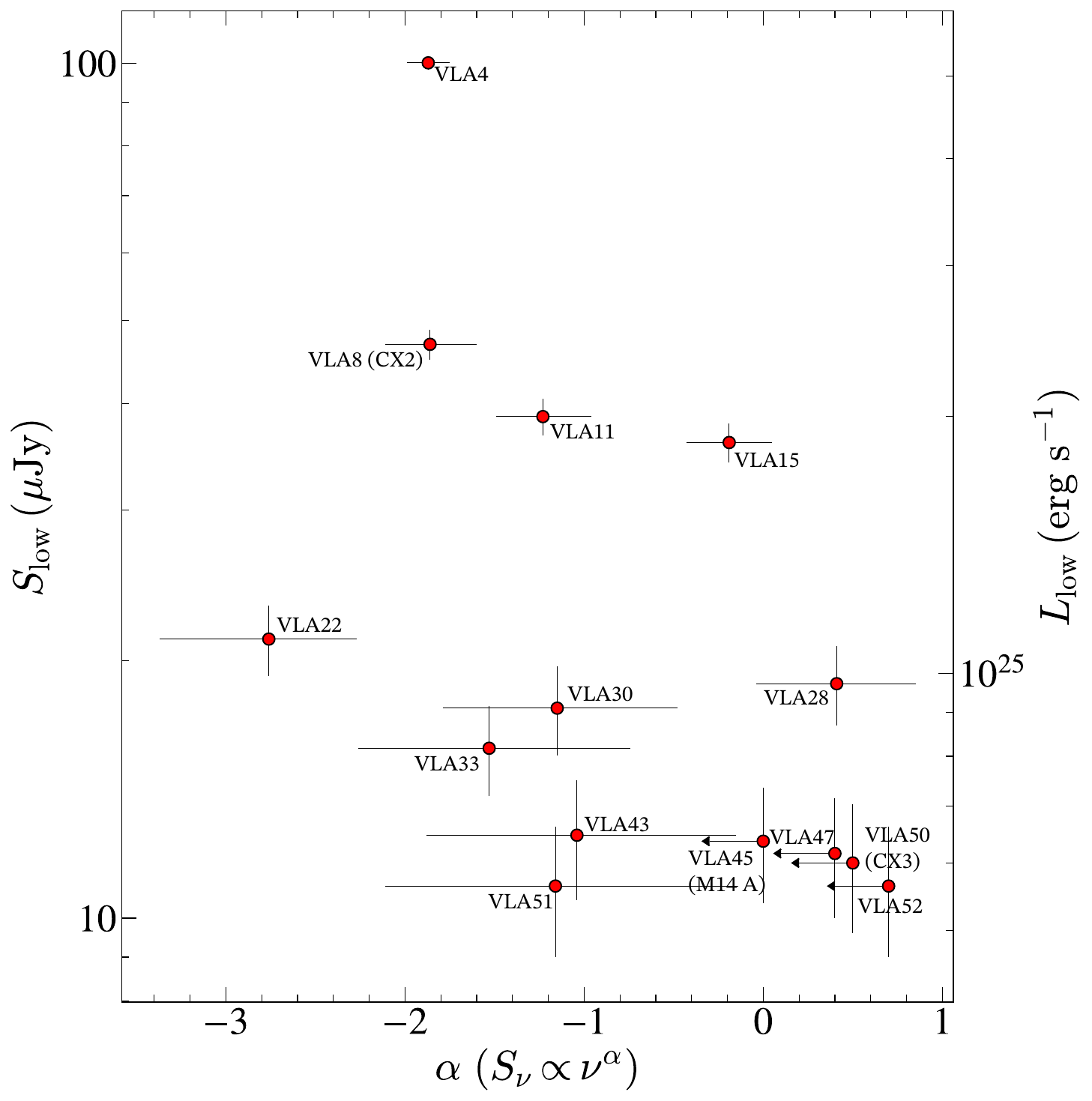}
    \caption{Flux density at the lower frequency sub-band ($\slow$) plotted against radio spectral indices ($\alpha$) for the \nradiosource\ radio sources within the cluster $\rh$. The error bars correspond $1\,\sigma$ uncertainties (see \citetalias{Shishkovsky20}). The right scale shows the corresponding radio luminosity at $\nulo$ assuming cluster distance of $9.3\,\kpc$ and flat spectra ($\alpha=0$).}
    \label{fig:radio-cmd}
\end{figure}

The processes of data reduction, imaging and source extraction are described in \citet[][Sh20, hereafter]{Shishkovsky20}; the resulting radio images at the low and high sub-band have noise levels of 1.8 and $1.7\,\mujyperbeam$ with synthesised beam sizes of $0\farcs46\times 0\farcs40$ and $0\farcs35\times 0\farcs29$, respectively. \citetalias{Shishkovsky20} reports a total of 14 compact radio sources at the $5\,\sigma$ level within the cluster half-light radius \citep[$\rh=1\farcm3$;][2010 edition]{Harris96}, of which 4 sources are in the core (Figure \ref{fig:x_ray_image}). In Figure \ref{fig:radio-cmd}, we plot the $\slow$ and $\alpha$ values for these sources.

\subsection{UV/optical/near-IR Observations}
\label{sec:hst-observations}
M14 was observed by the Wide-Field Camera 3 (WFC3) on board {\it the Hubble Space Telescope} (\hst) in August 2021 (GO-16283; PI D'Antona). Data were acquired with the UVIS imager with a range of UV, optical, and near-IR filters, including F275W ($\UV$), F336W ($\U$), F438W ($\B$), and F814W ($\I$). Individual exposures are arranged at intervals of 1--2 days from 2021-02-06 to 2021-02-13. For our analysis, we query individual calibrated \hst\ images (FLC files) from the Mikulski Archive for Space Telescopes (MAST) using the Hubble Search portal\footnote{\url{https://mast.stsci.edu/search/ui/\#/hst}}; these images have been flat-fielded and corrected for charge transfer inefficiency. A summary of observation is given in Table \ref{tab:hst_obs_summary}.

\begin{table*}
    \caption{A summary of \hst\ observations used in this work.}
    \begin{tabular}{ccccc}
    \toprule
     Proposal ID & Instrument & Filter & Number of exposures & Total exposure time (s) \\
    \midrule
     16283       & WFC3       & F275W ($\UV$) & 13 & 17172 \\
     16283       & WFC3       & F336W ($\U$)  & 10 & 7610  \\
     16283       & WFC3       & F438W ($\B$)  & 13 & 3760  \\
     16283       & WFC3       & F814W ($\I$)  & 11 & 1353  \\
    \bottomrule
    \end{tabular}
    \label{tab:hst_obs_summary}
\end{table*}

\section{Methodology}
\label{sec:methods}
\subsection{Imaging}
\subsubsection{UV/optical/near-IR}
To improve the signal-to-noise ratio of the images and help in the identification of faint counterparts to close binaries, we aligned and combined the FLC images for each filter. This was accomplished using the {\sc drizzlepac} Python package (version 3.5.1)\footnote{\url{https://www.stsci.edu/scientific-community/software/drizzlepac.html}}. The individual FLC images are firstly aligned to a reference frame (selected to be the FLC image with the longest exposure) by the {\tt tweakreg} tool, which are then ``drizzled" onto a common frame by  {\tt astrodrizzle}. To increase the image resolution and thus reduce potential crowding in the field, we set the combined image pixel scale to $0\farcs02$/pixel, which is half of the WFC3/UVIS scale.

\subsubsection{X-ray}
We filter and re-bin the processed event file to generate a $2\farcm72 \times 2\farcm72$ square %cutoff
image 
between $0.5$ and $10\,\kev$, which covers the whole $\rh$ of the cluster (Figure \ref{fig:x_ray_image}). Since the field is not crowded, we do not over-bin the event file and keep the pixels at their original sizes ($0\farcs5$). For further source detection processes, we also generate a fluxed image and an associated exposure map of the square %cutoff
image 
in the same energy band, using the {\sc ciao} {\tt fluximage} script.

\subsection{X-ray source detection}
We perform source detection with the {\sc ciao} {\tt wavdetect} tool on the X-ray %cutoff. 
image. 
To account for spatial variation of the source point spread function (PSF), we also include a PSF map of the region which is generated by the {\tt mkpsf} tool. These intermediate files are then given to the {\tt wavdetect} tool, which performs a wavelet transform on the input image at different scales and calculates the corresponding correlation coefficients for each pixel --- larger coefficients are considered more likely to belong to sources. We use scales of 1.0, 1.4, 2.0, 2.8 (powers of $\sqrt{2}$) and set the threshold significance to the invert of the image area ($9\times 10^{-6}$), so there will be approximately one false detection in the resulting source list\footnote{\url{https://cxc.cfa.harvard.edu/ciao/threads/wavdetect/}}.

\citet[][B20, hereafter]{Bahramian20} reported a total of 7 sources within $\rh$, including 5 that are deemed confident, one marginal detection, and one likely false detection. Our {\tt wavdetect} run found all of the 5 confident detections but missed the marginal and likely false detections. We check a $1\arcsec$ circular region around the marginal source, and found only 3 events between $0.5$ and $10\,\kev$. We therefore only include the five confident \citetalias{Bahramian20} detections in our catalogue. Beyond these 5 sources, our run also detected two faint sources that have $5$ and $4$ events within $1\arcsec$, and {\tt wavdetect} estimates 4.7 and 3.8 net counts for them, respectively. In fact, it is very hard to make an unambiguous conclusion on the genuineness of these very faint sources given the short exposure time, so we keep these two new sources in our catalogue. %based on a conventional net count limit of $4$ (need a ref here).

We sort the catalogue by the {\tt wavdetect}-estimated net counts in descending order and rename the sources by appending a number (starting from 1) to ``CX". The net counts are also used to calculate the $95\%$ error radii ($\perr$) of the sources according to the empirical formula in \citet{Hong05}, which are used in our processes of counterpart searching. These sources are listed in Table \ref{tab:x_ray_catalogue}.

% \begin{table*}
% \centering
% \caption{A catalogue of X-ray sources within $\rh$ of M14.}
%     \begin{tabular}{llcccccl}
%         \toprule
%         CX$^a$ & CXOU\_J$^b$ & RA (ICRS)$^c$ & DEC (ICRS)$^c$ & $\doff^d$ & Counts & $\perr^e$ & Comments \\
%            &         & (hh:mm:ss.ss)   & $^\circ$:$\arcmin$:$\arcsec$    & ($\arcmin$) & (0.5--10 keV) & ($\arcsec$) &  \\
%         \midrule
% 		1 & 173738.25-031441.7 & 17:37:38.24 & $-$03:14:41.85 & 0.54 & 32.3 & 0.36 & Nova Ophiuchus 1938 \\
% 		2 & 173734.32-031442.0 & 17:37:34.32 & $-$03:14:42.31 & 0.44 & 20.7 & 0.38 & VLA8; MSP? \\
% 		3 & 173737.39-031351.8 & 17:37:37.39 & $-$03:13:51.84 & 0.95 & 6.6 & 0.52 & VLA50; AGN? \\
% 		4 & 173736.04-031448.2 & 17:37:36.04 & $-$03:14:48.21 & 0.05 & 6.5 & 0.50 & SSG; AB? \\
% 		5 & 173736.03-031506.2 & 17:37:36.03 & $-$03:15:06.65 & 0.35 & 5.7 & 0.53 & AB? \\
% 		6 & \nodata   & 17:37:34.22 & $-$03:13:57.31 & 0.93 & 4.7 & 0.59 & \nodata \\
% 		7 & \nodata   & 17:37:38.07 & $-$03:15:50.40 & 1.19 & 3.7 & 0.70 & AGN? \\
%         \bottomrule
%         \multicolumn{8}{l}{$^a$New \chandra\ IDs; numbers ordered by net counts.}\\
%         \multicolumn{8}{l}{$^b$Source IDs from \citetalias{Bahramian20}.}\\
%         \multicolumn{8}{l}{$^c$Sky coordinates aligned to \gaia\ DR3.}\\
%         \multicolumn{8}{l}{$^d$Angular offsets from the cluster centre in $\arcmin$.}\\
%         \multicolumn{8}{l}{$^e$\citet{Hong05} 95\% error radii in $\arcsec$.}
%         \end{tabular}
%     \label{tab:x_ray_catalogue}
% \end{table*}

\begin{table*}
\centering
\caption{A catalogue of X-ray sources within $\rh$ of M14.}
    \begin{tabular}{llccccc}
        \toprule
        CX$^a$ & CXOU\_J$^b$ & RA (ICRS)$^c$ & DEC (ICRS)$^c$ & $\doff^d$ & Counts & $\perr^e$ \\
           &         & (hh:mm:ss.ss)   & $^\circ$:$\arcmin$:$\arcsec$    & ($\arcmin$) & (0.5--10 keV) & ($\arcsec$) \\
        \midrule
		1 & 173738.25-031441.7 & 17:37:38.24 & $-$03:14:41.85 & 0.54 & 32.3 & 0.36 \\
		2 & 173734.32-031442.0 & 17:37:34.32 & $-$03:14:42.31 & 0.44 & 20.7 & 0.38 \\
		3 & 173737.39-031351.8 & 17:37:37.39 & $-$03:13:51.84 & 0.95 & 6.6 & 0.52 \\
		4 & 173736.04-031448.2 & 17:37:36.04 & $-$03:14:48.21 & 0.05 & 6.5 & 0.50 \\
		5 & 173736.03-031506.2 & 17:37:36.03 & $-$03:15:06.65 & 0.35 & 5.7 & 0.53 \\
		6 & \nodata   & 17:37:34.22 & $-$03:13:57.31 & 0.93 & 4.7 & 0.59 \\
		7 & \nodata   & 17:37:38.07 & $-$03:15:50.40 & 1.19 & 3.7 & 0.70 \\
        \bottomrule
        \multicolumn{7}{l}{$^a$New \chandra\ IDs; numbers ordered by net counts.}\\
        \multicolumn{7}{l}{$^b$Source IDs from \citetalias{Bahramian20}.}\\
        \multicolumn{7}{l}{$^c$Sky coordinates aligned to \gaia\ DR3.}\\
        \multicolumn{7}{l}{$^d$Angular offsets from the cluster centre in $\arcmin$.}\\
        \multicolumn{7}{l}{$^e$\citet{Hong05} 95\% error radii in $\arcsec$.}
        \end{tabular}
    \label{tab:x_ray_catalogue}
\end{table*}

\subsection{X-ray variability}
\revise{We ran the {\sc ciao} {\tt glvary} script to check for source variability. {\tt glvary} separates the source events into multiple time bins and applies the Gregory-Loredo algorithm \citep{Gregory92} to detect significant variation between these bins. The script compute a variability index between $0$ and $10$, which considers indices greater than $6$ a confident sign of variability.

None of our X-ray sources are identified as variable. CX5 and CX6 have variability indices of 2 and 1, respectively, while all other sources have variability indices of 0. The results on the very faint sources (CX3--7) are not conclusive, considering the small number of counts (<10) available for the analysis.
}

\subsection{X-ray spectral analysis}
We noted that all M14 sources in the \citetalias{Bahramian20} catalogue are modelled with the hydrogen column density parameter ($\nh$) as a free parameter, with which one would get sensible constraints on $\nh$ for sources that have sufficient counting statistics. This, however, might not be optimal for faint sources. Since all of our sources have low counting statistics ($<40$ counts), we perform a separate spectral analysis with $\nh$ fixed to the cluster value $5.23\times 10^{21}\,\mathrm{cm^{-2}}$ (\citetalias{Bahramian20}). 

Spectra are extracted from the \chandra\ event file using the {\sc ciao} {\tt specextract} script, which are then regrouped to at least one count per bin using the {\tt dmgroup} tool. We perform fitting using the {\sc sherpa} software \citep[version 4.15.1;][]{Sherpa23} and use the W-statistics \citep[\wstat;][]{Cash79}. Since \chandra\ ACIS has a decreasing quantum efficiency at low energies\footnote{\url{https://cxc.cfa.harvard.edu/ciao/why/acisqecontamN0013.html}}, we ignore spectral channels below $0.5\,\kev$ for our fitting, and in all of our fitting, we use the {\tt wilm} abundance \citep{Wilm00} and the \citet{Verner96} cross section table.

We construct the fitting model by convolving the {\sc xspec} {\tt TBabs} absorption model with a selection of additive models, including (1) a power-law model (\pow), (2) a blackbody model (\bbody), and (3) a emission spectral model for diffuse plasma (\apec). Besides model normalisation, each of these additive component has one free parameter; for \pow\ this is the photon index ($\Gamma$) defined in $F_E \propto E^{-\Gamma}$, where $F_E$ is the specific energy flux at energy $E$; for \bbody\ and \apec, the other free parameter is the blackbody or plasma temperature in $kT$. {\sc sherpa} reports $\wstat$ and the associated ``Q-value" as a goodness of fit measure. The latter is the probability of observing the reduced statistics or a larger value assuming that the fit model is genuine. We choose the best-fit model to be the one that has the reduced \wstat\ closest to $1$ while having an acceptable Q-value ($\geq 5\%$); we report their maximum likelihood estimates and $1\,\sigma$ (68.27\%) confidence intervals on fit parameters in Table \ref{tab:spectral_fit_results}. We want to point out that fitting to CX6, and CX7's spectra have degrees of freedom (dof) less than 4, so they are too faint to get valid constraints on the parameters; we therefore apply a power-law model with a fixed $\Gamma=2$ to just fit the normalisation. 

\begin{figure}
    \centering
    \includegraphics[width=\columnwidth]{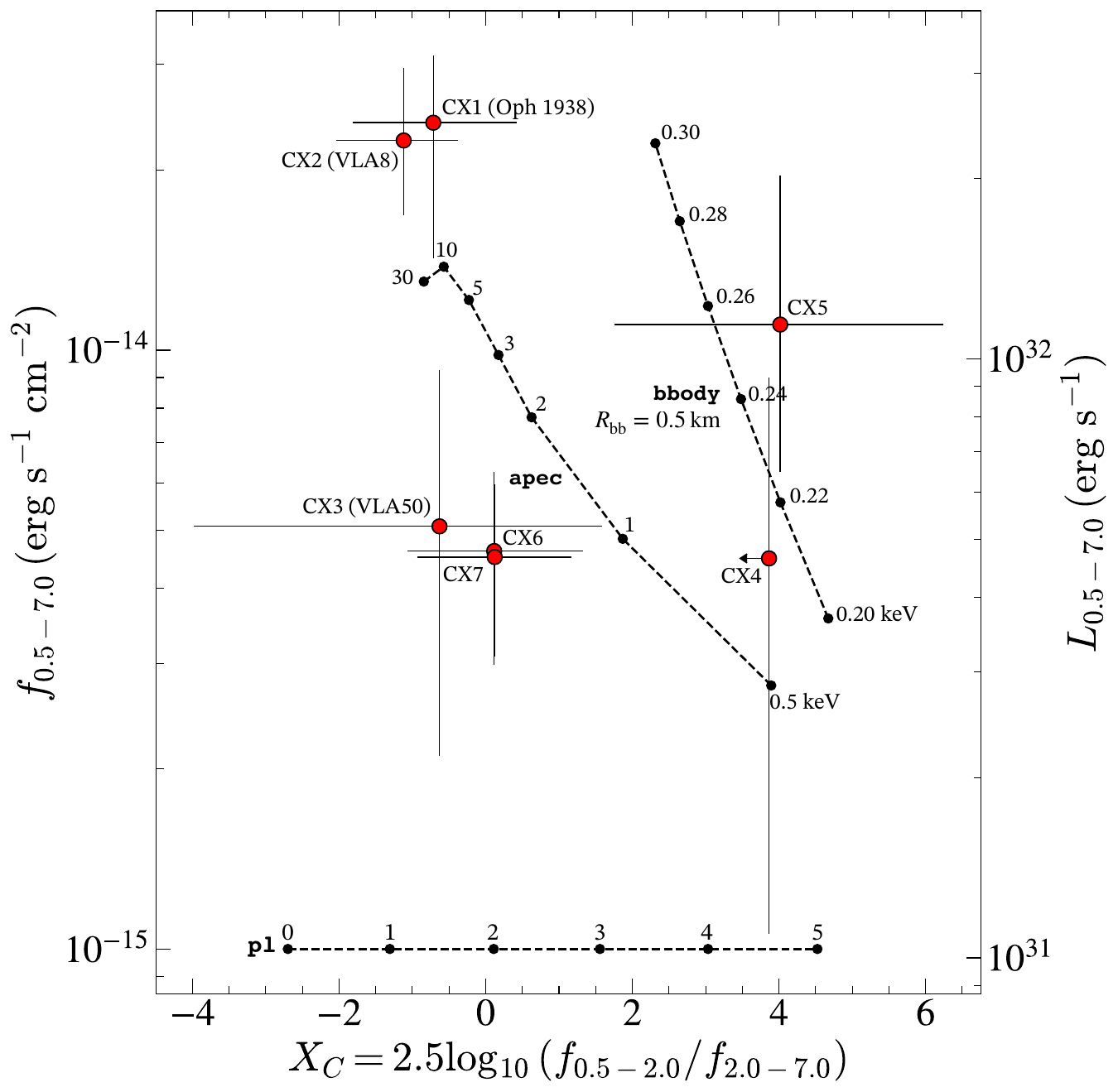}
    \caption{X-ray hardness ratio (eq. \ref{eq:x-ray-hardness}) vs. unabsorbed $0.5-7\,\kev$ fluxes for X-ray sources in Table \ref{tab:x_ray_catalogue}. Error bars are at the $1\,\sigma$ (68.27\%) level. Filled circles and dashed lines present hardness ratios and fluxes for \pow, \bbody, and \apec\ models at different parameters. For the \pow\ and \apec model, the normalisation is fixed at some arbitrary values; while for the \bbody\ model, we use a emitting region radius ($\rbb$) of $0.5\,\mathrm{km}$ to determine the normalisation.}
    \label{fig:x-ray_cmd}
\end{figure}

\begin{figure*}
    \centering
    \includegraphics[width=0.9\textwidth]{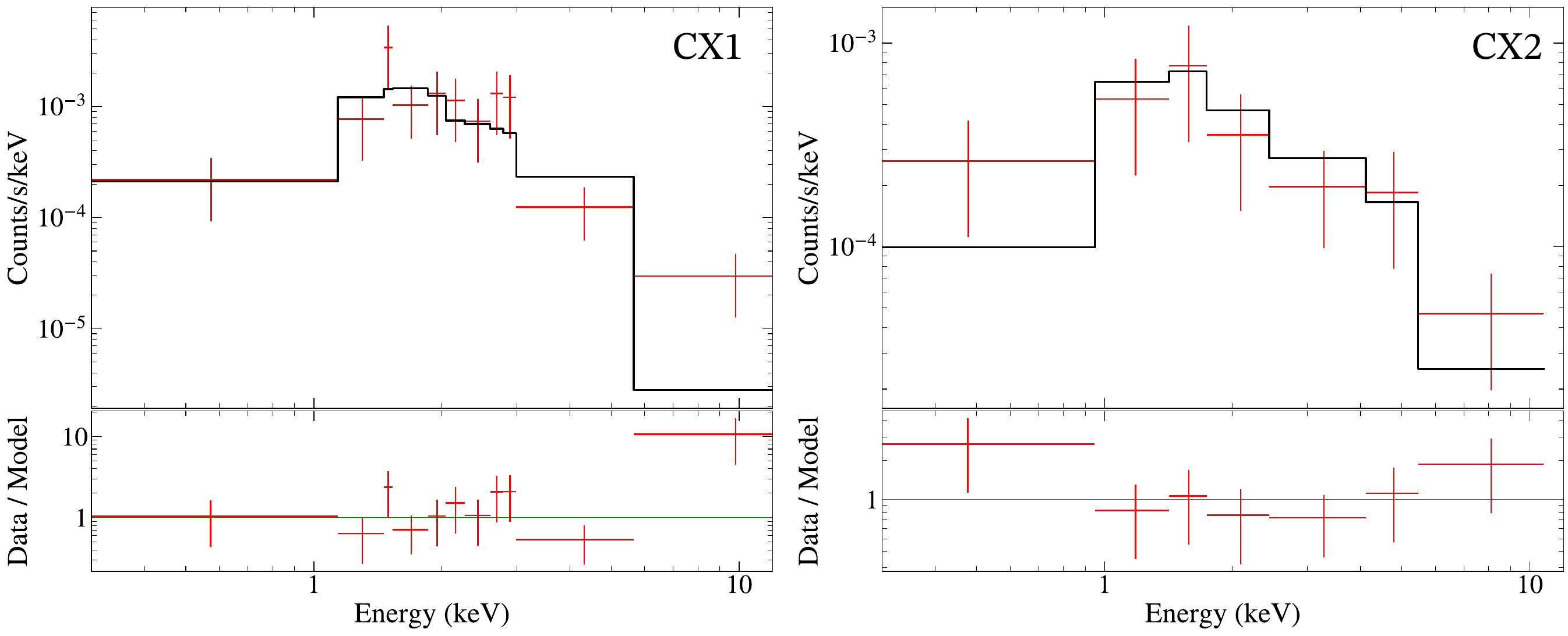}
    \caption{\chandra\ spectra (red) of CX1 (left) and CX2 (right) with the best-fit \bbody\ and \pow\ models (solid black lines) overplotted; the lower panels shows data to model ratios. Both spectra are re-grouped only for plotting purpose.}
    \label{fig:cx1_and_cx2_spectra}
\end{figure*}

With the best-fit model, we calculate fluxes using model parameters sampled from the best-fit parameters using the {\sc sherpa} {\tt sample\_energy\_flux} function; this is performed for a soft ($0.5-2\,\kev$), a hard ($2-7\,\kev$) X-ray, and a broad ($1-10\,\kev$) band, and the fluxes are denoted by $\fsoft$, $\fhard$, and $\fbroad$, respectively. $\fbroad$ is used to compare with radio fluxes (see Sec. \ref{sec:x-ray-to-radio-flux-ratio}), while the soft and hard bands are used to define the X-ray hardness ratio, $X_C$:
\begin{equation}
    X_C = 2.5 \log_\mathrm{10}\lrb{\frac{\fsoft}{\fhard}}.
    \label{eq:x-ray-hardness}
\end{equation}
In Figure \ref{fig:x-ray_cmd}, we plot the hardness and $\fsoft + \fhard$ fluxes for the X-ray sources\revise{, and in Figure \ref{fig:cx1_and_cx2_spectra}, we show the spectra and best-fit models for CX1 and CX2, the two relatively bright sources.}

% \begin{table*}
% \renewcommand{\arraystretch}{1.3}
% \centering
% \caption{Spectral fitting results of best-fit models. Errors are at the $1\,\sigma$ (68.27\%) level.}
%     \begin{tabular}{lcllccccc}
%     \toprule
%     CX &  Model & \multicolumn{2}{c}{Parameter} & $\fsoft$ & $\fhard$ & $\fbroad$ & \wstat\ (dof) & Q-value \\
%        &        &  name & value & \multicolumn{3}{c}{$(\times 10^{-15}\ergscm)$} & & \\
%     \midrule
%     1  & \bbody & $kT$     & $0.8\pm 0.1$ & $8.25^{+8.36}_{-5.86}$ & $16.87^{+32.23}_{-13.86}$ & $23.80^{+40.89}_{-18.70}$ & 43.2 (28) & 0.03  \\
%     2  & \pow   & $\Gamma$ & $1.1\pm 0.4$ & $5.86^{+2.63}_{-2.55}$ & $17.21^{+19.83}_{-9.55}$ & $30.61^{+36.29}_{-17.28}$ & 17.95 (19) & 0.52 \\
%     3  & \pow   & $\Gamma$ & $1.1^{+1.0}_{-0.9}$ & $1.92^{+1.77}_{-1.73}$ & $3.55^{+15.82}_{-3.40}$ & $6.44^{+34.10}_{-6.04}$ & 7.00 (6) & 0.32 \\
%     4  & \pow   & $\Gamma$ & $0.7^{+1.0}_{-0.9}$ & $<3.12$ & $2.38^{+11.11}_{-2.15}$ & $4.44^{+20.13}_{-3.76}$ & 4.38 (5) & 0.50 \\
%     5  & \pow   & $\Gamma$ & $4.6^{+1.2}_{-1.1}$ & $10.64^{+9.05}_{-4.72}$ & $0.25^{+0.90}_{-0.19}$ & $1.78^{+1.86}_{-0.84}$ & 5.96 (4) & 0.24 \\
%     6  & \pow   & $\Gamma$ & $2.0^\dag$ & $2.44^{+1.23}_{-1.21}$  & $2.18^{+1.10}_{-1.09}$ & $3.98^{+2.06}_{-2.01}$ & 5.59 (5) & 0.31 \\
%     7  & \pow   & $\Gamma$ & $2.0^\dag$ & $2.40^{+1.10}_{-1.07}$ & $2.16\pm 0.99$ & $3.93^{+1.76}_{-1.82}$ & 2.89 (3) & 0.40 \\
%     \bottomrule
%     \end{tabular}
%     \label{tab:spectral_fit_results}
% \end{table*}

\begin{table*}
\renewcommand{\arraystretch}{1.3}
\centering
\caption{Spectral fitting results of best-fit models. Errors are at the $1\,\sigma$ (68.27\%) level.}
    \begin{tabular}{lcllccccc}
    \toprule
    CX &  Model & \multicolumn{2}{c}{Parameter} & $\fsoft$ & $\fhard$ & $\fbroad$ & \wstat\ (dof) & Q-value \\
       &        &  name & value & \multicolumn{3}{c}{$(\times 10^{-15}\ergscm)$} & & \\
    \midrule
       & \pow & $\Gamma$ & $1.3^{+0.3}_{-0.3}$ & $11.30^{+3.09}_{-3.01}$ & $24.75^{+8.11}_{-6.05}$ & $43.59^{+15.06}_{-10.46}$ & 48.27 (28) & 0.01 \\
      CX1 & \apec & $kT$ & $>5.71\, \kev$ & $12.40^{+2.35}_{-2.48}$ & $22.62^{+4.69}_{-4.82}$ & $37.89^{+7.88}_{-7.93}$ & 47.74 (28) & 0.01\\
	  & \bbody & $kT$ & $0.8^{+0.1}_{-0.1}\, \kev$ & $8.18^{+1.97}_{-3.94}$ & $15.79^{+6.81}_{-8.93}$ & $22.73^{+8.06}_{-11.20}$ & 43.24 (28) & 0.03 \\
   \midrule
	 CX2 & {\tt pl} & $\Gamma$ & $1.1^{+0.4}_{-0.4}$ & $5.89^{+2.15}_{-2.31}$ & $16.50^{+6.85}_{-5.10}$ & $29.92^{+13.60}_{-9.43}$ & 17.95 (19) & 0.53 \\
	 CX3 & {\tt pl} & $\Gamma$ & $1.1^{+1.0}_{-0.9}$ & $1.82^{+1.51}_{-1.64}$ & $3.26^{+3.89}_{-2.48}$ & $5.74^{+7.43}_{-3.81}$ & 7.00 (6) & 0.32 \\
	 CX4 & {\tt pl} & $\Gamma$ & $0.8^{+1.0}_{-0.9}$ & $<2.19$ & $3.27^{+4.39}_{-3.20}$ & $5.99^{+9.26}_{-5.45}$ & 3.13 (5) & 0.68 \\
	 CX5 & {\tt pl} & $\Gamma$ & $4.6^{+1.2}_{-1.1}$ & $10.77^{+8.49}_{-4.76}$ & $0.27^{+0.93}_{-0.20}$ & $1.80^{+1.98}_{-0.88}$ & 5.96 (4) & 0.20 \\
	 CX6 & {\tt pl} & $\Gamma$ & $2.0^\dag$ & $2.43^{+1.25}_{-1.21}$ & $2.19^{+1.08}_{-1.11}$ & $4.02^{+2.01}_{-2.04}$ & 6.00 (5) & 0.31 \\
	 CX7 & {\tt pl} & $\Gamma$ & $2.0^\dag$ & $2.38^{+1.07}_{-1.07}$ & $2.13^{+0.99}_{-0.95}$ & $3.93^{+1.82}_{-1.77}$ & 2.89 (3) & 0.41 \\
    \bottomrule
    \multicolumn{9}{l}{$^\dag$ parameters frozen to this value during the fit.} \\
    \multicolumn{9}{l}{Fluxes have been corrected for cluster absorption.}
    \end{tabular}
    \label{tab:spectral_fit_results}
\end{table*}

\subsection{Astrometry}
Relative shifts between observations made by different instruments can be contributed by different calibration standards and/or cluster proper motion between epochs. For better positional identification, we choose the \gaia\ DR3 astrometry (epoch=2016.0) as the reference frame and inspect its alignment with \chandra, \vla, and \hst\ astrometry. The radio observation was made only $\approx 0.4\,\mathrm{yr}$ earlier than the \gaia\ epoch, which corresponds to only minor shifts in RA and DEC of $\approx -1.4\,\mas$ and $\approx -2.0\,\mas$ adopting cluster proper motion in \citet{Vasiliev21}. \chandra\ and \hst\ observations are $7.5$ and $5.7$ years apart from the \gaia\ epoch, so proper motion would contribute shifts $\approx 0\farcs04$, which %has an unignorable 
may have a detectable 
effect on our identification considering the pixel scale of the \hst\ images is 0\farcs02. We therefore perform separate astrometric alignment for them.

\subsubsection{UV/optical/near-IR}
We align the drizzle-combined images to \gaia\ using source positions in the \gaia\ Data Release 3 (DR3). We query the DR3 database for sources within a radius of $\rh$ and select sources that have precise positions. For $\UV$, $\U$, and $\B$, we cut \gaia\ sources that have RA and DEC uncertainties $\geq 0.1\,\mas$; for $\I$, we select relatively fainter (\gaia\ G-band magnitude $\geq$ 19) sources as brighter sources are affected by saturation in the $\I$ image; as a result, we loosen the upper limit on positional uncertainty to $1\,\mas$ to maintain a sensible source number. We use {\tt DAOStarFinder} module in {\sc photutils} \citep[version 1.6.0;][]{Bradley23} to find sources in the \hst\ images. {\tt DAOStarFinder} applies the {\tt DAOFIND} algorithm that was developed as part of the {\sc daophot} \citep{Stetson87} photometry software. The source positions are then matched up to the \gaia\ positions to find relative offsets (\gaia $-$\hst) in RA and DEC, which are summarised in Table \ref{tab:hst-astrometry}.

\begin{table}
    \caption{\gaia$-$\hst\ offsets in RA and DEC. Uncertainties are at the $1\,\sigma$ level.}
    \centering
    \begin{tabular}{lccc}
    \toprule
    Filter & $\Delta$RA  & $\Delta$DEC & $N_\mathrm{match}^a$ \\
           & ($\arcsec$) & ($\arcsec$) &   \\ 
    \midrule
    $\UV$  & $-0.044\pm 0.004$ & $-0.046\pm 0.005$ & 312 \\
    $\U$   & $-0.066\pm 0.005$ & $-0.007\pm 0.006$ & 312 \\
    $\B$   & $-0.047\pm 0.004$ & $-0.027\pm 0.005$ & 312 \\
    $\I$   & $-0.106\pm 0.009$ & $-0.005\pm 0.010$ & 164 \\
    \bottomrule
    \multicolumn{4}{l}{$^a$Number of \gaia-\hst\ matches used.}
    \end{tabular}
    \label{tab:hst-astrometry}
\end{table}

\subsubsection{X-ray}
\label{sec:x-ray-astrometry}
\chandra\ ACIS-S has an absolute astrometric uncertainty of $\approx 0\farcs5$\footnote{\url{https://cxc.harvard.edu/cal/ASPECT/celmon/\#catalogs}}, and since our {\tt wavdetect} sources are all relatively faint ($\lesssim 50$ counts), this leads to non-negligible offsets relative to the \gaia\ frame. Therefore, confident matches are needed for correcting the \chandra\ boresight. The known pulsar PSR J1737$-$0314A (M14 A, hereafter) has a well-established timing position \citep{Pan21}, but it is reported by \citet{Pan21} as 
a black widow pulsar that hosts an extremely low-mass companion --- very difficult to detect with optical observations. Instead, we refine the boresight correction of the \chandra\ catalogue in a iterative manner.

\begin{figure}
    \centering
    \includegraphics[width=\columnwidth]{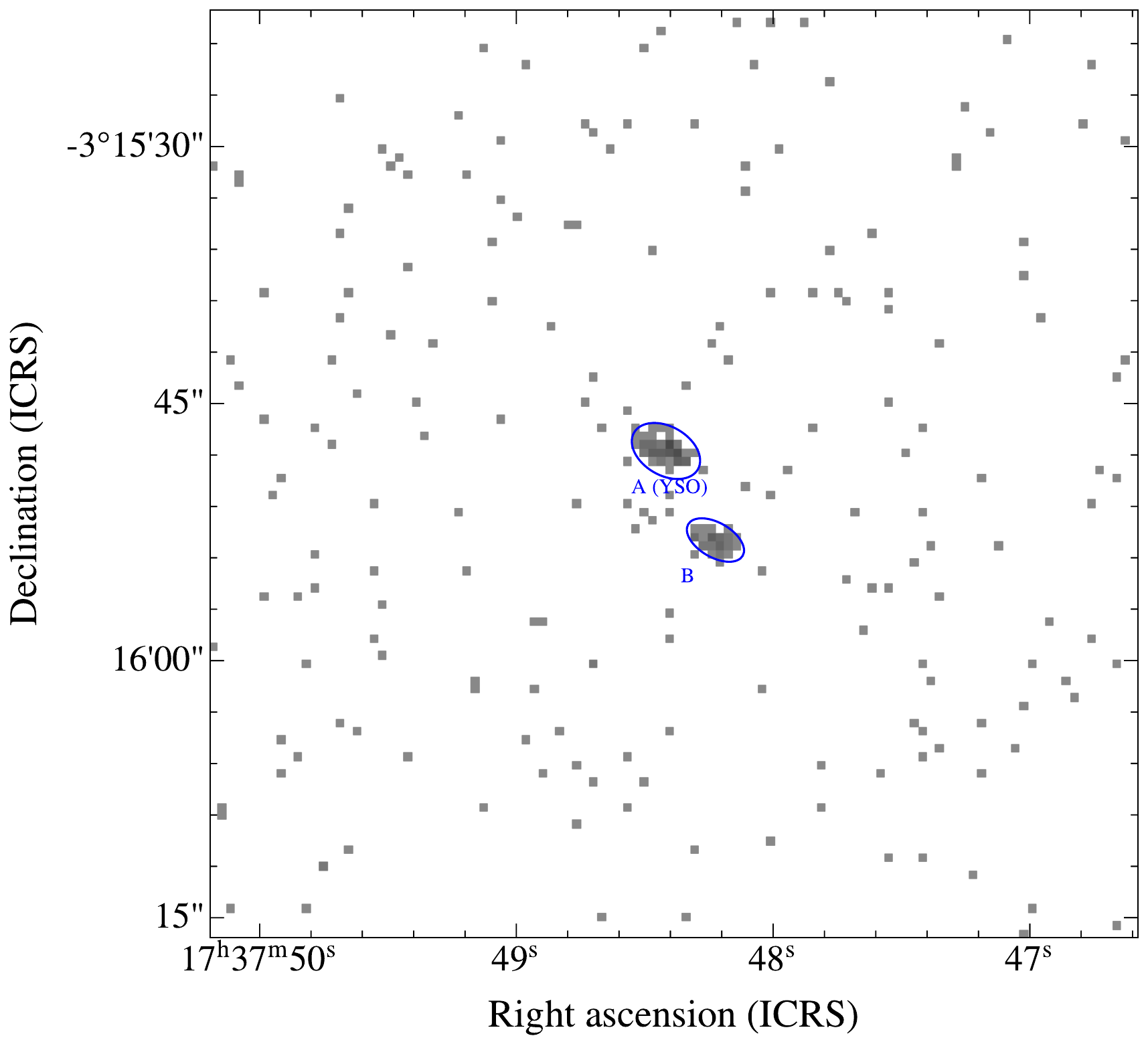}
    \caption{A $0\farcm9\times 0\farcm9$ X-ray ($0.5-10\,\kev$) image including the two off-axis sources A (a YSO) and B; north is up and east is to the left. The blue ellipses are regions used to calculate centroids of the sources.}
    \label{fig:chandra_offaxis_image}
\end{figure}

We first apply a rough correction using off-axis sources. There are two relatively bright \chandra\ sources that are outside $\rh \approx 3\farcm2$ southeast to M14; their PSFs are elongated because of their large off-axis angles; we denote the brighter north source and fainter south source with A and B (Figure \ref{fig:chandra_offaxis_image}), respectively. We then manually encircle source A and B with a $2\farcs1\times 1\farcs4$ and a $1\farcs8\times 1\farcs0$ elliptical region and use the {\sc ciao} {\tt dmstat} tool to get the centroid positions of them. For each of A and B, we found one close \gaia\ source within $\approx 0\farcs2$, both of which have parallaxes and proper motions suggestive of foreground stars. In fact, both sources have proper motions that are $\approx 3\,\sigma$ away from the cluster proper motion in the vector point diagram (VPD; Figure \ref{fig:gaia_vpd}). The \gaia\ counterpart to A ({\tt source\_id}=4368928767444987648) was also identified with the known young stellar object (YSO) UCAC4 434$-$071758 \citep{Zari18}; The \gaia\ counterpart to B ({\tt source\_id}=4368928767444986880) has a $\bp-\rp$ colour of $0.7$ (de-reddened adopting the \gaia\ {\tt ebpminrp\_gspphot}), suggestive of a yellow dwarf (see Table \ref{tab:offaxis_sources} for a summary of source A and B). The \gaia\ astrometry of this source yield a satisfactory single-star solution (with renormalised unit weight error, or {\tt ruwe}=1.05) so it is not likely a binary. This source is bright enough to also have a BP/RP spectrum that exhibits a likely broad H$\alpha$ absorption feature (Figure \ref{fig:bp-rp-spectrum}); this could be a result of rotational broadening, which could induce strong coronal activity and contribute to X-ray emission. We therefore also consider the \gaia\ counterpart to source B genuine. With the \gaia\ and centroid positions for A and B, we align \chandra\ to \gaia\ using the averaged offset in RA ($0\farcs14$) and DEC ($-0\farcs19$).

These corrected coordinates are then used to search for possible optical counterparts, and as a result, we found confident \hst\ counterparts to CX1 and CX2 (see Sec \ref{sec:individual-sources}). We therefore refine the \chandra\ boresight with the mean offset between these two sources and their counterparts. This gives (\gaia$-${\tt wavdetect}) $\Delta \mathrm{RA}=-0\farcs15$ and $\Delta \mathrm{DEC}=-0\farcs13$.

\subsection{UV/optical/near-IR photometry}
\label{sec:optical-uv-photometry}
Detailed processes of UV/optical/near-IR photometry have been presented in \citet[][D22, hereafter]{DAntona22}, including correction for differential reddening. The \citetalias{DAntona22} photometry catalogue includes magnitudes in the WFC3 filters and the number of epochs where the magnitude is well-measured, with which we can further reduce the catalogue to make colour-magnitude diagrams (CMDs). Specifically, for each pair of filters, we use a least conservative condition to include sources that have at least one good measurement in both filters. In Figure \ref{fig:hst-counterparts-cmd}, we present the $\UV-\U$, $\U-\B$, and $\B-\I$ CMDs for M14. These CMDs are further used in identifying X-ray and radio sources (Sec \ref{sec:counterpart-search}). 

\subsection{Counterpart searches}
\label{sec:counterpart-search}
Searching for UV/optical/near-IR counterparts to X-ray or radio sources is based on positional matching; however, even though \chandra\ and \vla\ sources are localised to sub-arcsecond scales, the high number density of UV/optical/near-IR sources means that chance coincidences could confuse the genuineness of the matches. For some sources, it is possible to reduce the degeneracy by complementing the results with UV/optical/near-IR photometric properties which can be associated with relevant astrophysical processes. For example, UV photometry has been broadly used in identifying CVs as they commonly exhibit UV excesses that are attributed to emission from the hot WD surface and/or ongoing accretion, while CVs are more consistent with the main sequence in optical CMDs \citep[e.g.,][]{Pooley02a,Edmonds03,Zhao19, Cohn21}. Most BY Draconis (BY Dra) type of ABs are found to be consistent with the binary sequence (i.e., slightly above the main sequence) and often show H$\alpha$ emission \citep[e.g.,][]{Cohn10,Pallanca17,Lugger23}. In essence, counterparts to X-ray emitting close binaries are expected to be photometric outliers, and the relative scarcity of these outliers also significantly reduces their probability of being chance coincidences.

For \chandra\ sources, we cross-match X-ray source positions with the \citetalias{DAntona22} catalogue, searching for counterparts within  $\perr$. These counterparts are then further checked with the CMDs for photometric outliers. For the \vla\ sources, we use source-specific search radii (denoted with $\perrradio$) that are set to the %maximum between
larger of 
$0.1$ of the synthesised beam size at $\nulo$ and the positional uncertainty reported in \citetalias{Shishkovsky20}. Since \vla\ sources have error circles $\approx 10$ times smaller than the \chandra\ sources, position matches are much less likely to be chance coincidences. Finally, we also cross-match the \chandra\ and \vla\ catalogues to find potential associations; the search radii must account for X-ray and radio positional uncertainties, so we use $\sqrt{\perr^2 + \perrradio^2}$ to calculate them.

In addition, we also compare the radio timing position of M14 A \citep{Pan21} with source positions in the \chandra, \vla, and \citetalias{DAntona22} catalogues. The search results are further discussed in Sec \ref{sec:individual-sources}.

\subsection{X-ray/optical flux ratio}
\begin{figure}
    \centering
    \includegraphics[width=\columnwidth]{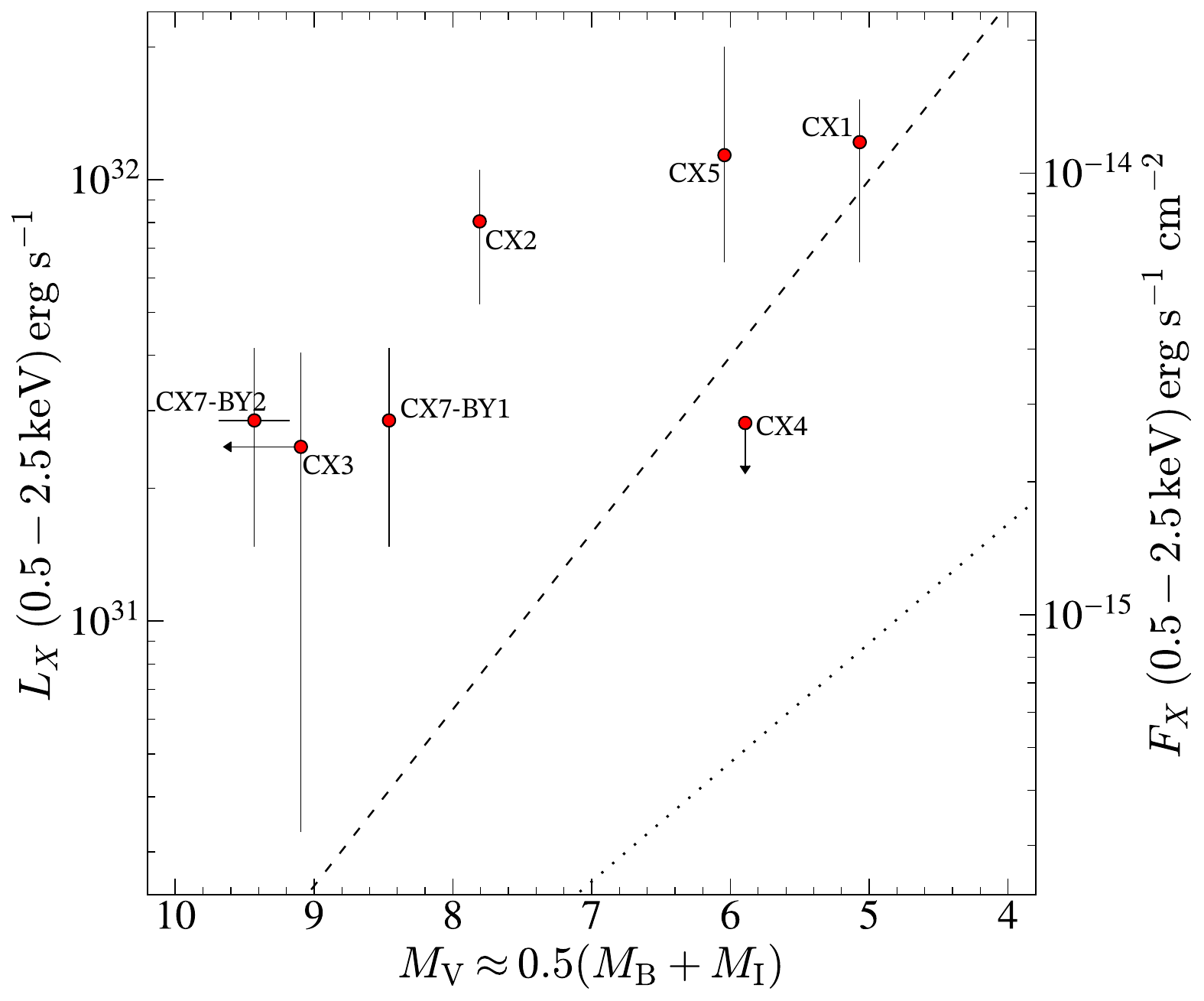}
    \caption{$0.5-2.5\,\kev$ luminosity vs. approximated V-band absolute magnitude for X-ray sources that have \hst\ counterparts. The dashed line is the empirical separatrix separating CVs above from ABs below, and the dotted line marks the empirical X-ray luminosity upper limits for stars and known ABs \citep{Verbunt08}.}
    \label{fig:x-ray-to-optical-ratio}
\end{figure}

\revise{One helpful indicator for source identification is the X-ray/optical flux ratio. For X-ray sources with optical counterparts, one can compare their X-ray/optical ratios to empirical relations that separate different source classes. We follow the band and filter combination that has been conventionally used in other works \citep[e.g.,][]{Edmonds03, Bassa04, Verbunt08}, calculating $0.5-2.5\,\kev$ X-ray luminosity and V-band absolute magnitude ($\mv$) using the cluster distance from \citet[][]{Harris96} ($d=9.3\,\kpc$). As there is no V-band observation, we approximate $\mv$ with the averaged $\I$ and $\B$ absolute magnitudes ($\mb$ and $\mi$), i.e., $\mv\approx 0.5(\mb + \mi)$. Sources with \hst\ counterparts are plotted in Figure \ref{fig:x-ray-to-optical-ratio}. We also include an empirical line separating cluster CVs from ABs and a line that marks upper limits of X-ray luminosities for nearby stars and ABs \citep{Verbunt08}.
}

\subsection{X-ray/radio flux ratio}
\label{sec:x-ray-to-radio-flux-ratio}
\begin{figure*}
    \centering
    \includegraphics[width=0.8\textwidth]{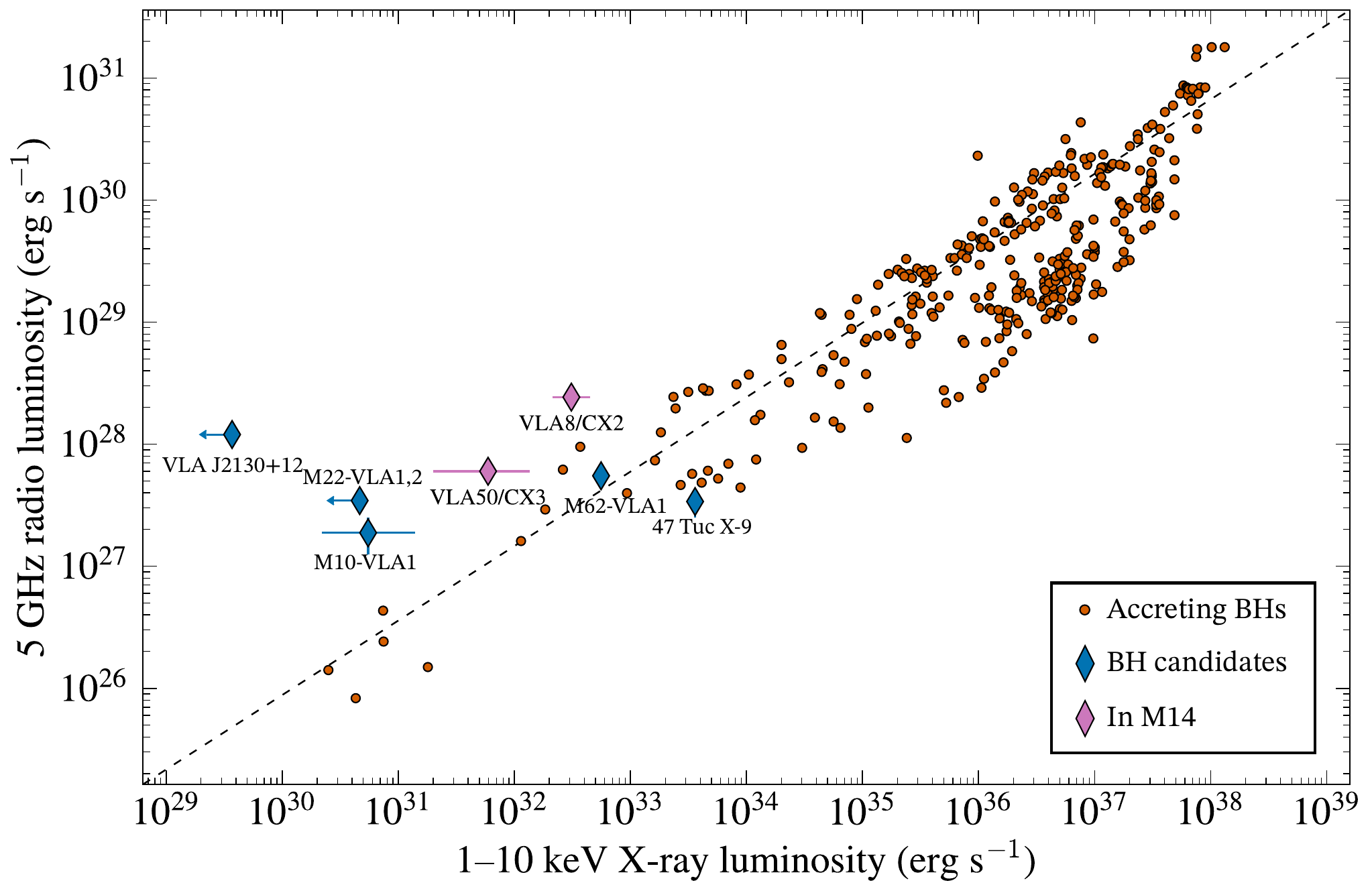}
    \caption{$5\,\ghz$ radio vs. $1-10\,\kev$ X-ray luminosities plotted for accreting BHs (filled orange circles) and BH candidates (filled blue diamonds) from \citet[][and references therein]{Bahramian22}; radio luminosities are calculated assuming flat spectra ($\alpha=0$). Two X-ray sources in M14 that have \vla\ counterparts, namely CX2/VLA8 and CX3/VLA50, are indicated with filled magenta diamonds. The dashed line shows the X-ray-radio correlation for accreting BHs from \citet{Gallo14}.}
    \label{fig:x-ray-to-radio-flux-ratio}
\end{figure*}

\revise{X-ray sources that have radio counterparts can be further checked against other accreting compact objects for their X-ray/radio ratios. Typically, it has been shown that accreting BHs exhibit a tight correlation between their X-ray and radio luminosities \citep[e.g.,][]{Gallo14}. In Figure \ref{fig:x-ray-to-radio-flux-ratio}, we plot $\nulo$ luminosity against $1-10\,\kev$ luminosity for CX2 and CX3, together with a compilation of accreting BHs and BH candidates from \citet{Bahramian22}. }

\subsection{UV/optical/near-IR variability}
\begin{figure*}
    \centering
    \includegraphics[width=0.9\textwidth]{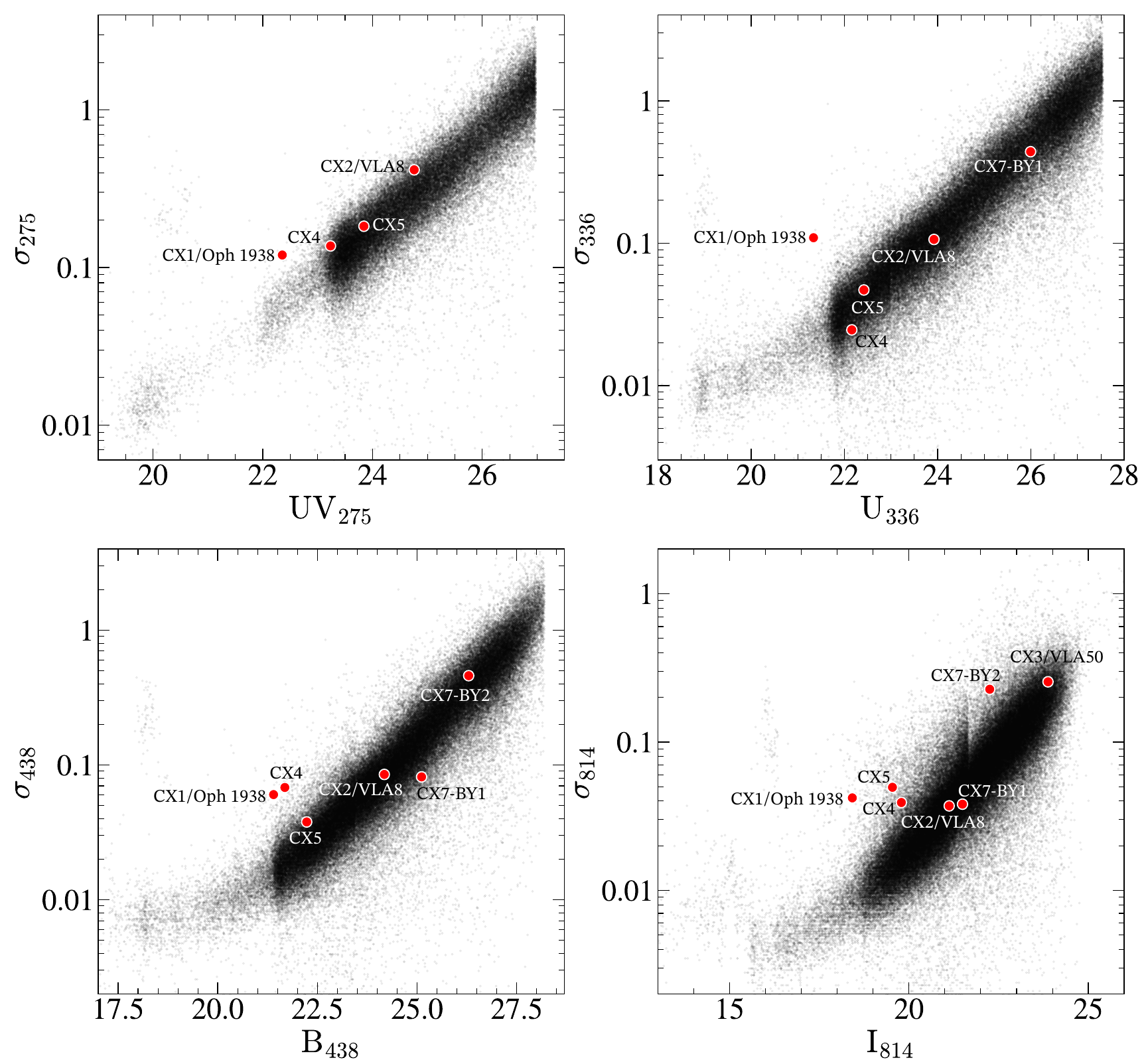}
    \caption{Magnitude RMS plotted against magnitude for 4 \hst\ filters. Counterparts of interest are overplotted with different markers. Sources with an excess of RMS compared to the bulk of points have variability.}
    \label{fig:rms-mag-plot}
\end{figure*}

In addition to photometric positions on CMDs, variability information could also be important to break the degeneracy. In practice, CVs are expected to show strong variability in the UV and optical bands \citep[e.g.,][]{Warner03,RiveraSandoval18, Lugger17}, and ABs %can show stochastic flares that could also render them variables. --I haven't seen this in photometry
may show variation due to eclipses, ellipsoidal variability, and/or starspots \citep{Albrow01,Lugger23}.
MSP companions may also show variability produced by ellipsoidal variations, or by varying visibility of a heated face of the companion \citep{Callanan95,Orosz03}.
Since the \citetalias{DAntona22} catalogue contains epoch photometry, an effective way of checking for variability is to search for excesses in root mean square (RMS; denoted by $\sigma$ with a subscript indicative of the filter name) magnitude, over the bulk of non-variables at a given magnitude. In Figure \ref{fig:rms-mag-plot}, we plot RMS magnitude vs. magnitude for four \hst\ filters, where strong variables tend to lie above the bulk of points in the plot.
% \subsubsection{Cluster membership}

% \subsection{Radio-to-X-ray}

\subsection{Chance coincidence}
\label{sec:chance-coincidence}
\begin{figure*}
    \centering
    \includegraphics[width=0.9\textwidth]{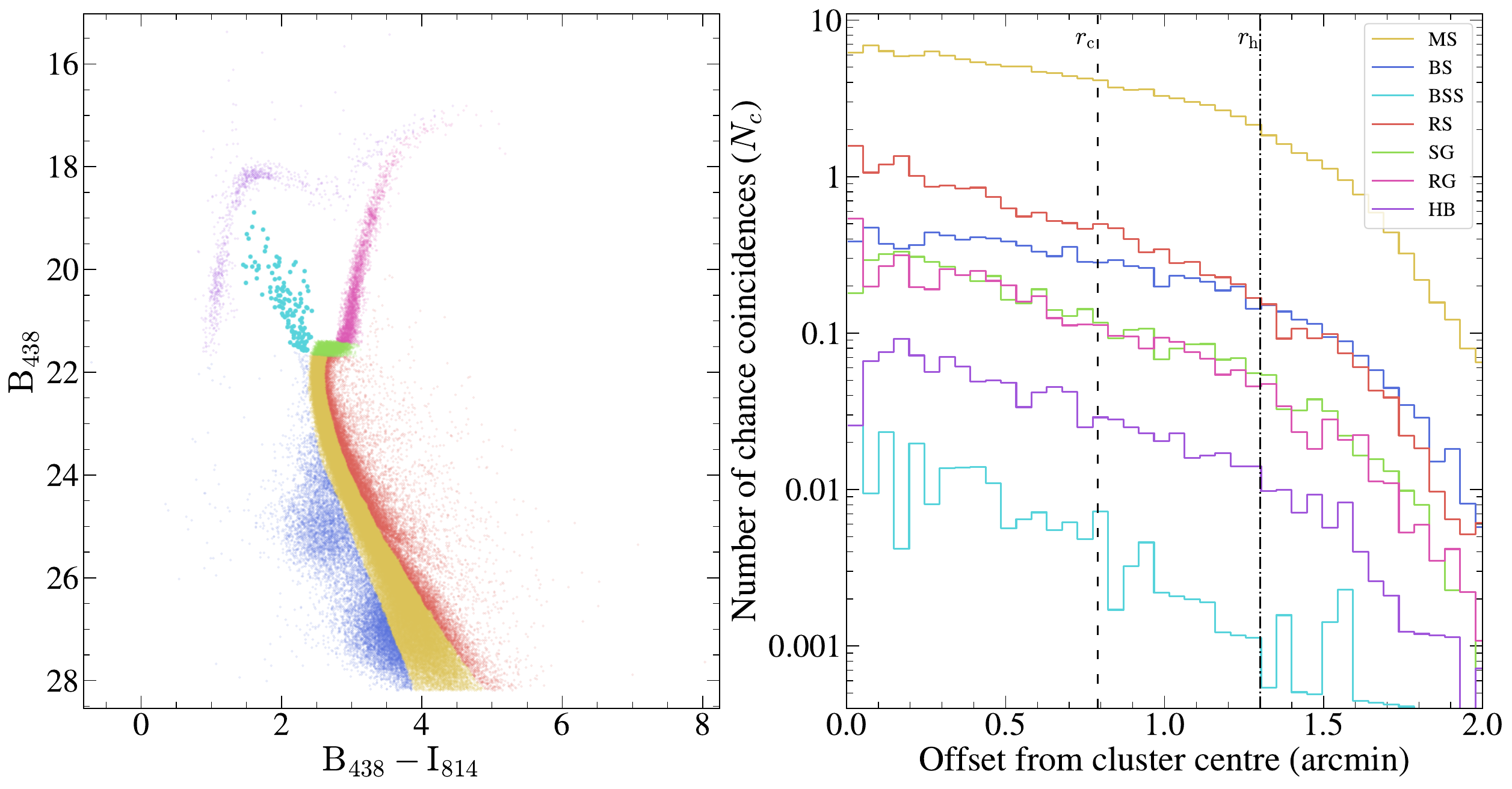}
    \caption{{\it Left}: $\BIB$ CMD showing different subpopulations (colours online). {\it Right}: Number of chance coincidences ($\nc$) as a function of offset from the cluster centre assuming a search radius of $0\farcs5$; the vertical dashed and dashed-dotted lines mark the core ($\rc$) and half-light radius ($\rh$), respectively. Shorthands: MS: main sequence, BS: sources with blue excesses, BSS: blue stragglers, RS: sources with red excesses, SG: subgiants, RG: red giants.}
    \label{fig:chance-coincidence-numbers}
\end{figure*}

The number of predicted chance coincidences, denoted with $\nc$, is calculated following the methods in \citet{Zhao20b}. In short, we divide up sources in the $\B-\I$ CMD into subgroups by applying polygonal selection regions using the {\sc glueviz} software \citep{Beaumont15, Robitaille17}. The sources are then divided into those on the main sequence (MS), subgiant branch (SG), red giant branch (RG),  horizontal branch (HB), blue stragglers (BSS), and sources with blue (BS) and red (RS) excesses (Figure \ref{fig:chance-coincidence-numbers}). The number densities (counts per projected area) of sources in these subgroups are tallied for different radial offsets from the GC centre, and the $\nc$ with a X-ray/radio source is then simply the product of the relevant number density and the area of the corresponding search regions. In Figure \ref{fig:chance-coincidence-numbers}, we show $\nc$ values calculated with the average error radius ($\approx 0\farcs5$) for different subpopulations.

\section{Results and discussions}
\label{sec:results}
\begin{figure*}
    \centering
    \includegraphics[width=0.8\textwidth]{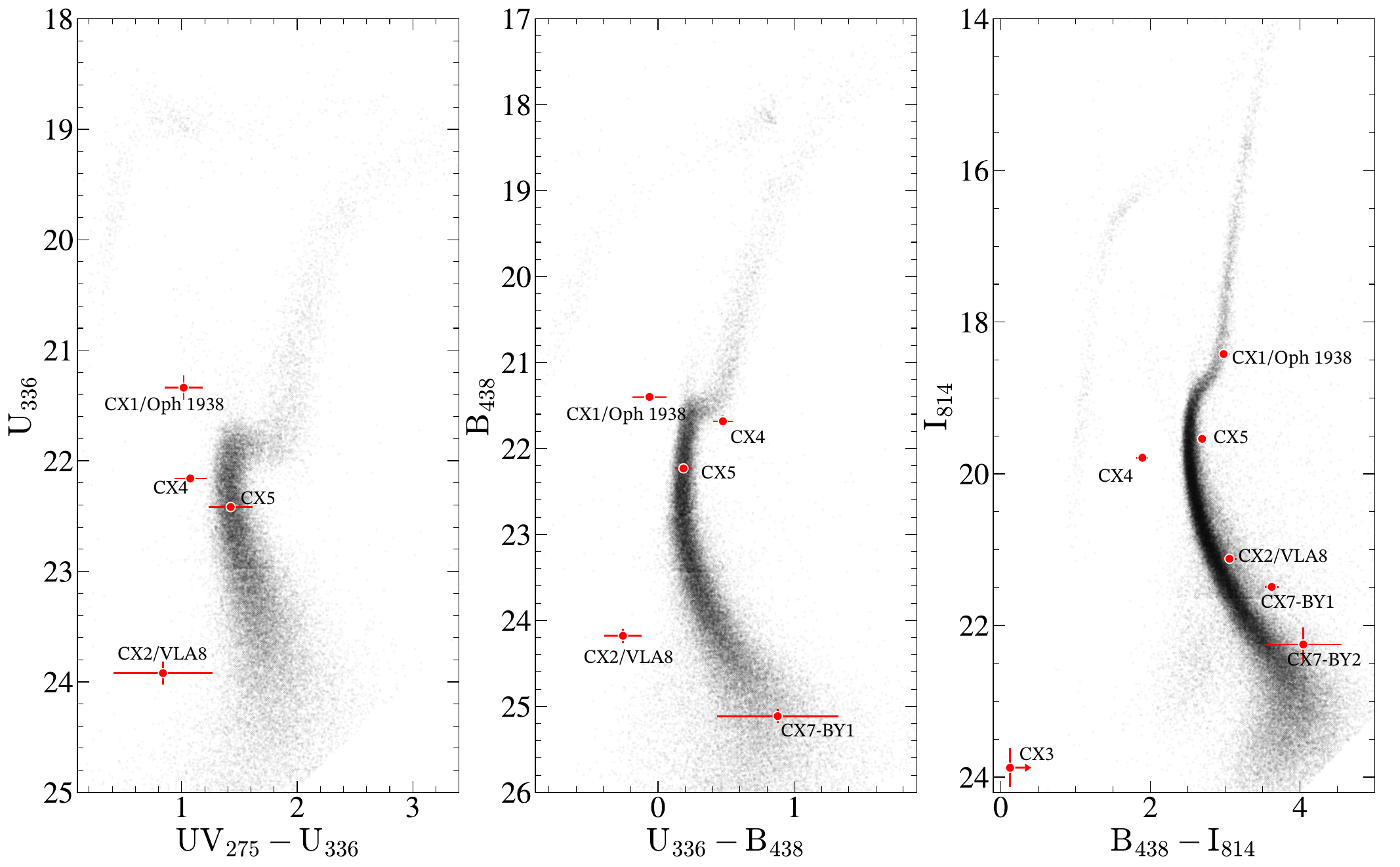}
    \caption{$\UVUU$, $\UBB$, and $\BII$ CMDs of $M14$, with counterparts to \chandra\ and \vla\ sources overplotted.}
    \label{fig:hst-counterparts-cmd}
\end{figure*}

\begin{figure*}
    \centering
    \includegraphics[width=0.8\textwidth]{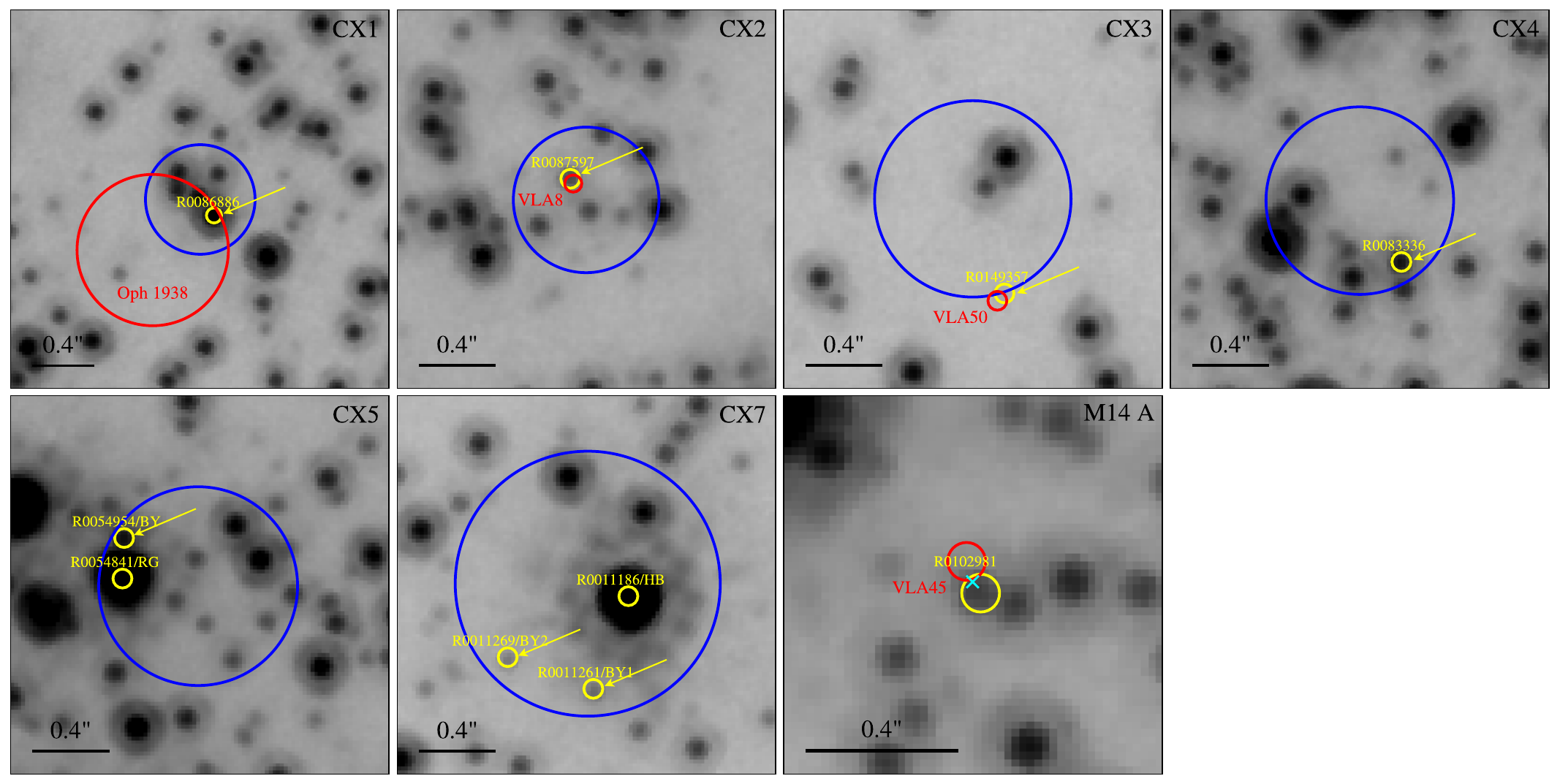}
    \caption{$\I$ finding charts of \chandra\ and \vla\ sources and their \hst\ counterparts (colours online). North is up and east is to the left. The chart for CX1 is $2\farcs5 \times 2\farcs5$ and $2\arcsec\times 2\arcsec$ in size for the other ``CX" sources; the chart for M14 A is $1\arcsec\times 1\arcsec$ in size. The blue circle in each chart indicates the \citet{Hong05} $95\%$ error circle (radius of $\perr$), red circle is the \vla\ error circle (radius of $\perrradio$), and yellow circles ($0\farcs05$ in radius) marks the \hst\ sources that are mentioned in the discussion (Sec \ref{sec:individual-sources}); the most likely counterparts are indicated by arrows. The cyan cross in the M14 A chart indicates the timing position from \citet{Pan21} of M14 A after being backtracked with cluster proper motion, and the red circle close to CX1 marks the position of the classic nova candidate Oph 1938 advanced to 2016 by the cluster proper motion.}
    \label{fig:hst-finders}
\end{figure*}

As a brief summary of results, cross-matching our \chandra\ catalogue with the \vla\ catalogue shows that CX2 matches the position of VLA8, while CX3 marginally matches with VLA50. The known pulsar M14 A's timing position is consistent with VLA45 (Table \ref{tab:radio-counterparts}). By comparing magnitude and colours with the bulk photometry, we find potential \hst\ counterparts to $\ncounterpartstox$ of the X-ray sources. A cross-match of these \hst\ counterparts with the \gaia\ DR3 catalogue does not reveal any match, so no further proper motion information is available for them. Details are summarised in Table \ref{tab:hst_counterparts} for \hst\ counterparts; in Figure \ref{fig:hst-counterparts-cmd}, we present photometry of the \hst\ counterparts; and in Figure \ref{fig:hst-finders}, we presents the $\I$ finding charts for sources with \hst\ counterparts. In the following sections, we present discussions on individual sources.

\begin{table*}
\renewcommand{\arraystretch}{1.2}
\centering
\caption{Summary of \hst\ counterparts.}
\resizebox{\textwidth}{!}{
    \begin{tabular}{llcccccclll}
    \toprule
    \multicolumn{2}{c}{ID} & \multicolumn{4}{c}{Photometry} & $\nc$ & $\probagn^a$ & \multicolumn{3}{c}{Comments} \\
    CX & \hst & $\UV$ & $\U$ & $\B$ & $\I$ & & & Colour & Variability & Identification \\
    \midrule
    1 & R0086886 & $22.36\pm 0.12$ & $21.34\pm 0.11$ & $21.40\pm 0.06$ & $18.42\pm 0.04$ & $0.01$ & $0.42$ & UV excess & All bands & Oph 1938 \\
     2 & R0087597 & $24.76\pm 0.42$ & $23.92\pm 0.11$ & $24.18\pm 0.09$ & $21.12\pm 0.04$ & $0.41$ & $0.31$ & UV excess & No & VLA8; MSP? \\
     3 & R0149357 & \nodata & \nodata & \nodata & $23.88\pm 0.25$ & $3.57$ & $0.82$ & Faint; only in $\I$ & No & VLA50; AGN? \\
     4 & R0083336 & $23.24\pm 0.14$ & $22.16\pm 0.02$ & $21.68\pm 0.07$ & $19.79\pm 0.04$ & $1.24$ & $0.00$ & SSG & in $\B$ and $\I$ & RS CVn or MSP? \\
     5 & R0054954 & $23.84\pm 0.18$ & $22.42\pm 0.05$ & $22.23\pm 0.04$ & $19.54\pm 0.05$ & $0.85$ & $0.21$ & Binary sequence & only in $\I$ & BY Dra? \\
     7-BY1 & R0011261 & \nodata & $25.99\pm 0.44$ & $25.11\pm 0.08$ & $21.49\pm 0.04$ & $0.23$ & $0.93$ & Binary sequence & only in $\I$ & BY Dra? \\
     7-BY2 & R0011269 & \nodata & \nodata & $26.29\pm 0.46$ & $22.25\pm 0.23$ & $0.23$ & $0.93$ & Binary sequence & only in $\I$ & BY Dra? \\
    \bottomrule
    \multicolumn{11}{l}{$^a$Probability of finding at least one AGN within the source's radial offset.}
    \end{tabular}
    }
    \label{tab:hst_counterparts}
\end{table*}

\begin{table}
    \centering
    \caption{Summary of \vla\ counterparts taken from \citetalias{Shishkovsky20}.}
    \begin{tabular}{llccc}
    \toprule
    \vla\ ID & Assoc. & $\slow$ & $\shigh$ & $\alpha$   \\
             &                 & \multicolumn{2}{c}{$(\mujy)$} & \\ 
    \midrule
    VLA8     & CX2             & $46.9\pm 1.9$ & $26.6\pm 1.8$ & $-1.86^{+0.25}_{-0.26}$ \\
    VLA45    & M14 A           & $12.3\pm 1.9$ & $<5.1$ & $<0.0$ \\
    VLA50    & CX3             & $11.6\pm 2.0$ & $<5.4$ & $<0.5$ \\
    \bottomrule
    \end{tabular}
    \label{tab:radio-counterparts}
\end{table}

\subsection{Individual sources}
\label{sec:individual-sources}
\subsubsection{CX1: association with Nova Ophiuchi 1938}
CX1's error circle is consistent with the source R0086886 that exhibits a strong UV excess in the $\UVUU$ and $\UBB$ CMDs (Figure \ref{fig:hst-counterparts-cmd}). In contrast, it appears near the base of the red giant branch when viewed in the $\BIB$ CMD. It is close to the BSS population in the $\UVUU$ CMD, and \revise{if it is indeed a BSS}, one expects a $\nc \approx 0.005$. If viewed as a RG the (Sec \ref{sec:chance-coincidence}), $\nc \approx 0.09$. \revise{Even if R0086886 is not a BSS or RG, its exotic colours still makes it} very unlikely to be a chance coincidence. This counterpart also shows marked variability in all four \hst\ bands (Figure \ref{fig:hst-counterparts-cmd}). This strong variability, however, is unlikely to fully account for the colour differences between CMDs, which points to an interpretation that the UV excess and red giant reflect two components of a binary.  

CX1 is close ($\approx 0\farcs5$) to the classic nova candidate Nova Ophiuchi 1938 (hereafter Oph 1938). This nova was first recorded on photographic plates at the David Dunlap Observatory in 1938 and identified 
by \citet{SawyerHogg64}. \citet{Shara89} obtained ground-based CCD observations of Oph 1938's field and obtained the epoch-1938 position down to $\approx 1\arcsec$ using the positions of nearby bright stars; this coordinate, however, was later found by \citet{Margon91} to be at odds with refined observations with the \hst/Faint Object Camera (FOC), from which the Oph 1938 position was re-measured with an uncertainty of $0\farcs5$. We convert this position to the ICRS frame and bring it to the \gaia\ epoch by applying the cluster proper motion in \citet{Vasiliev21}, giving ICRS RA (2016)=17:37:38.26(3), DEC (2016)=$-$03:14:42.2(5). The resulting error circle of Oph 1938 is consistent with both CX1 and R0086886 (Figure \ref{fig:hst-finders}). We therefore argue that both CX1 and R0086886 are counterparts to Oph 1938, making it the second classic nova recovered in a Galactic globular cluster, after 
Nova T Scorpii in M80 \citep{Dieball10}.

None of the additive models provide a satisfactory fit to CX1's X-ray spectrum; the best %of all
is a \bbody\ model with a temperature of $ 0.8\pm 0.1\,\kev$ and an emitting region of $84^{+24}_{-19}\,\mathrm{m}$. This model gives an X-ray luminosity ($0.5-7.0\,\kev$) of $\approx 2.4\times 10^{32}\,\ergps$, placing CX1 in the hard regime of faint GC X-ray sources. Despite the poor quality of the fit, the X-ray hardness is consistent with well-identified CVs in other GCs \citep[e.g.,][]{Heinke05}. The UV excess and strong variability in all bands could be attributed to %residual accretion onto the white dwarf surface; 
substantial mass transfer through an accretion disc; \revise{a CV nature is also favoured by its X-ray/optical ratio (Figure \ref{fig:x-ray-to-optical-ratio}),} which could be further confirmed by emission features (e.g., H$\alpha$) with future spectroscopic or photometric studies. A relatively high rate of mass transfer may be expected in a CV which has experienced a nova eruption \citep{Warner03}. Studying CVs nearly 100 years after a nova explosion can be helpful in understanding the relationship between nova explosions and mass transfer rates \citep[e.g.][]{Shara17}. 

Our measured X-ray luminosity can provide a lower limit (dependent on the assumed white dwarf compactness, i.e. its mass) to the mass transfer rate in this system. \referee{Extrapolating the best-fit model flux (Table \ref{tab:spectral_fit_results}) to between $0.002$--$25\,\kev$ with the WebPIMMS tool\footnote{\url{https://heasarc.gsfc.nasa.gov/cgi-bin/Tools/w3pimms/w3pimms.pl}}, and approximating the accretion luminosity with this flux; we can get an estimate on the lower limit of the mass accretion rate $\approx 7.0\times 10^{-11}\,\msunyr$ for a $0.5\,\Msun$ white dwarf.}

%Searching around this nominal position with a radius of $0\farcs2$, we find two faint sources: R0087305 and R0086981, both of which only have $\B$ and $\I$ magnitude available, but the latter shows a mild red excess while the former is consistent with the MS scatters at the faint end of the $\BIB$ CMD. R0086981 also appears to be a moderate variable in $\I$. Since variability is a expected feature of a classical novae \citep[e.g.,][]{Strope10}, R0086981 is more likely the true counterpart, but more detailed spectroscopic studies is needed to finally confirm its genuineness. 

\subsubsection{CX2/VLA8: a new MSP candidate?}
%Checking around 
In CX2's \chandra\ error circle, %we noted that 
the source R0087597 falls to the blue side of the $\UVUU$ and $\UBB$ CMDs but lies \referee{slightly to the red edge of the MS} in the $\BIB$ CMD (Figure \ref{fig:hst-counterparts-cmd}), corresponding to a late K dwarf \citep{Pecaut13}. The blue excess might have been an effect of strong variability, but there is no sign of variability for R0087597 in all \hst\ bands (Figure \ref{fig:rms-mag-plot}). The $\nc$ with MS sources at the offset of CX2 is around $3$, but if considered a BS source, its $\nc$ value is reduced to around $0.2$. Indeed, this could still be an overestimate for a source with a clear blue excess, considering a number of marginally blue (i.e., just slightly bluer than the MS) sources have been included in our estimate for $\nc$ (Figure \ref{fig:chance-coincidence-numbers}). R0087597 is also consistent with the steep radio source VLA8 (Figure \ref{fig:hst-finders}), and the small radio error circle renders a chance coincidence even more unlikely --- $\nc=0.003$. We therefore consider R0087597 a genuine counterpart to CX2 and VLA8. 

VLA8 has a well-constrained steep $\alpha=-1.86^{+0.25}_{-0.26}$, consistent with either a radio pulsar or an AGN \citep{Gordon21,Kramer98}, but its position within the cluster core makes it more likely to be a cluster member. Furthermore, the X-ray spectrum of CX2 can be fit by a relatively hard power-law ($\Gamma=1.1\pm 0.4$) at a $0.5-7.0\,\kev$ luminosity of $\approx 2.3\times 10^{32}\,\ergps$. The hardness and luminosity overlap ABs and faint CVs, but the former is strongly argued against by the high X-ray/optical ratio (Figure \ref{fig:x-ray-to-optical-ratio}). 

\referee{CX2/VLA8's blue and red colour combination makes a CV interpretation plausible, where the blue colours are mostly contributed by an accretion disc and/or a white dwarf, while the $\B-\I$ could be from a bloated donor. VLA8's bright radio luminosity ($L_\mathrm{low} = 2.4\times 10^{28}\,\mathrm{\ergps}$; Figure \ref{fig:x-ray-to-radio-flux-ratio}), however, could be an mild counterargument against a canonical CV (see e.g., \citealt{Ridder23} for a comprehensive summary of radio and X-ray measurements of CVs). Some more unusual magnetic CVs like AE Aquarii \citep{Bookbinder87}, LAMOST J024048.51+195226.9 \citep{Thorstensen20, Garnavich21, Pelisoli22}, and AR Sco \citep{Marsh16}, could be brighter during flares, but not as steep as VLA8 \citep[e.g.,][]{Pretorius21}. 

The steep radio spectral index suggests an MSP nature. However, in an MSP the optical emission arises only from the companion, which makes it difficult to generate the observed unusual colours. A white dwarf companion for an MSP could generate a UV excess \citep[e.g.,][]{Rivera-Sandoval15}, or a redback MSP with a late-type, low-mass companion could produce red optical colours \citep[e.g.,][]{Ferraro01}, but an MSP will not have both. Although irradiation of parts of a redback can make some parts hotter than others, examination of redbacks in globular cluster CMDs does not reveal similar examples of such different colours \citep[e.g.,][]{Ferraro01, Edmonds02, Pallanca13}. A more plausible interpretation is that CX2/VLA8 is a transitional MSP \citep[tMSP; ][]{Archibald09, Hill11, deMartino13, Papitto13, Linares14} that switches between an accretion-powered LMXB (at the \chandra\ and \hst\ observations) and a rotation-powered radio pulsar phase (when the \vla\ observation was performed), considering that its colour combination is similar to other LMXBs observed in GCs \citep[e.g., ][]{Edmonds02}. Despite the type of the binary MSP, these scenarios are all challenged by the lack of variability in the counterpart (Figure \ref{fig:rms-mag-plot}), while orbital modulation is often observed in their optical light curves \citep[e.g.,][]{Breton13, Papitto18}. In all, we leave CX2/VLA8's nature unsettled until more robust evidence of a pulsar (e.g., radio pulsations) and/or of membership (e.g., via proper motion) is available.}

% The UV excesses could be from a white dwarf companion \citep[e.g.,][]{Rivera-Sandoval15}, but this is hard to reconcile with the peculiar colour combination. Another possibility is a redback MSP scenario, where a late-type, low-mass ($\approx 0.1-0.5\,\msun$) companion is continuously irradiated by pulsar emission. Indeed, there are two new redback MSPs newly discovered by \citet{Pan21} that have no timing position available yet (also see discussion in Sec \ref{sec:other-msps-in-m14}), and one of which could be CX2. However, irradiation alone is insufficient to account for the strong UV excess.

% due to its lack of blue/UV excesses. If CX2 is indeed a pulsar, this luminosity and spectrum would suggest a relatively bright redback type of binary MSP \citep{Roberts14}, and interaction between the pulsar wind and companion stellar wind would then be responsible for the hard and non-thermal X-ray spectrum. The UV excess might be produced by pulsar irradiation of one side of the companion. Membership information (e.g., via proper motion analysis) is essential to provide more robust confirmation of CX2's nature.

\subsubsection{CX3/VLA50}
CX3's error circle is marginally consistent with that of VLA50; the latter contains one faint \hst\ source R0149357 that has photometry only in the $\I$ band (Figure \ref{fig:hst-finders}) with no sign of variability (Figure \ref{fig:rms-mag-plot}). \revise{If one adopts a $\B$ magnitude of ($\approx 24$) as a rough completeness limit, this gives a $\B-\I$ colour $\geq 0.12$, which overlaps with counterparts to various source classes} (Figure \ref{fig:hst-counterparts-cmd}). The probability of a spurious match between the \chandra\ and \vla\ sources is very small, as is the probability of a spurious \vla/\hst\ match, so we consider R0149357 to be robust. From the radio perspective, VLA50 is only detected at $\nulo$, giving an upper limit of $\alpha\leq 0.5$. \revise{The X-ray spectra can be fit by a hard power-law ($\Gamma=1\pm 1$); this, combined with its faint counterpart, makes CX3's X-ray/optical ratio too high to be an AB. CX3's X-ray and radio luminosities place it consistent with other accreting BHs in the X-ray-radio luminosity plot (Figure \ref{fig:x-ray-to-radio-flux-ratio}), and its $\alpha$ also overlaps with the flat-to-inverted ($\alpha \geq 0$) spectra observed in accreting BHs \citep[e.g.,][]{Plotkin19}. This might suggest an accreting BH nature; however, though poorly constrained, the X-ray photon index ($\Gamma=1\pm 1$) is only marginally consistent with quiescent BHs \citep[$\Gamma\approx 2$; e.g.,][]{Reynolds14}. Moreover, CX3's large offset from the cluster centre makes a background AGN contamination more %significance
likely.}
In all, the limited information on CX3 and VLA5 leads to various interpretations, although its location outside the core makes an AGN nature more likely than a member.

\subsubsection{CX4}
%We find three potential counterparts within the \chandra\ error circle of CX4. 
% There is a bright evolved \hst\ source (R0083345/CX4-RG) consistent with the RG of all CMDs (Figure \ref{fig:hst-counterparts-cmd}), which does not show variability in all \hst\ filters. RG counterparts are typical for RS CVn type of ABs (where of the components is a RG or SG star), but the $\nc$ value with a RG source is 0.3. A Poisson probability of encountering with at least one such RG source while expecting 0.3 is $0.26$, so chance coincidence is likely. The second counterpart, R0083345 (or CX4-BY), appears consistent with the binary sequence (i.e., slightly above the MS) in the $\BIB$ CMD (Figure \ref{fig:hst-counterparts-cmd}). This could indicate an AB nature and categorise CX4 as a BY Draconis type of AB (where both components are MS stars). However, if one treats this source as a RS, it has an $\nc$ value of 1.2, which again disfavours a genuine match with CX4, despite that R0083345 exhibits a very mild $\I$ variability (Figure \ref{fig:rms-mag-plot}). but it also makes chance coincidences more likely.

The fact that CX4 is only $3\arcsec$ from the cluster centre makes it more likely a cluster member. We noted that the source R0083336 has a varying colour across different CMDs. It exhibits a blue excess in the $\UVUU$ and $\BIB$ CMD; however, it appears as a sub-sub giant (SSG) in the $\UBB$ CMD (Figure \ref{fig:hst-counterparts-cmd}). SSGs form a rare population below the subgiant branch and brighter than the binary sequence sources. Such unusual loci on CMDs are hard to reconcile with standard single-star evolutionary models \citep{Leiner22}, and indeed, SSGs are found typically associated with binaries, and a fraction of them are also X-ray sources \citep{Geller17}. To get an estimate on the $\nc$ value for SSGs, we follow the loci of SSGs shown in \citet{Geller17} and select SSGs from the RS sub-population by applying criteria of $\B-\I\geq 3.0$ and $21.7 \leq \B \leq 22.5$. We found a total of 223 SSGs, and the corresponding $\nc$ value with a SSG source is only $1.8\times 10^{-4}$ even at CX4's vicinity to the cluster centre. Moreover, R0083336 shows marked $\B$ and moderate $\I$ variability (Figure \ref{fig:rms-mag-plot}), making it even too exotic to be a chance coincidence. We therefore conclude that this SSG is a genuine match.

\revise{The soft band flux is not constrained due to a dearth of counts, rendering CX4 a relatively hard source among the others. Its X-ray hardness can overlap with both CVs and ABs, but the X-ray/optical ratio argues against a CV nature (Figure \ref{fig:x-ray-to-optical-ratio})}; we hereby identify CX4 as a candidate AB.

\subsubsection{CX5}
There are two \hst\ sources worth noticing within CX5's error circle. R0054841 is a bright RG source in all three CMDs, which might be reminiscent of a RS Canum Venaticorum (RS CVn) type of AB, but it shows no sign of variability in any of the four \hst\ filters. Moreover, the $\nc$ value at CX5's radial offset is $\approx 0.27$, so this RG source could also be a chance coincidence. A more likely counterpart is R0054954/CX5-BY, located north of CX5-RG (Figure \ref{fig:hst-finders}). This source shows a mild red excess in the $\BIB$ CMD (Figure \ref{fig:hst-counterparts-cmd}) and has a clear variability in $\I$; its location is close to the SSGs in the $\BIB$ CMD, but the $\B-\I$ colour is less red. If we treat CX5-BY as a source on the binary sequence, CX5 could then be a BY Draconis type of AB. The $\nc$ value with a RS source for CX5 is $1$, suggesting that this match is a chance coincidence, but a RS source with  variability like CX5-BY is rarer. Though limited to the low counting statistics, the X-ray spectrum of CX5 can be fit with a soft \pow\ model ($\Gamma = 4.6^{+1.1}_{-1.2}$), which is reminiscent of faint CVs and ABs in GCs that tend to be softer \citep[e.g.,][]{Heinke05}. \revise{An AB interpretation, however, is mildly argued against by its X-ray/optical ratio (Figure \ref{fig:x-ray-to-optical-ratio})}. In all, we suggest that R0054954 is more likely the true counterpart and classify CX5 as a likely AB.

\subsubsection{CX7}
The error circle of CX7 contains several photometrically rare sources, including a HB and two RS sources. The HB source (R0011186) appears close to the red clump and has no variability in any \hst\ filters. Although CX7's error circle has a low $\nc$ value with a HB source ($\approx 0.03$) that argues against a chance coincidence, this is not sufficient to identify it with CX7. One of the two RS sources, R0011261/CX7-BY1, exhibits a marked red excess on the $\BIB$ CMD but is consistent with the MS on the $\UBB$ CMD. The other RS source, R0011269/CX7-BY2, lies at the fainter end of the $\BIB$ CMD and only has a moderate red excess; it is also a mild variable in $\I$. $\nc$ value with a RS at the offset of CX7 is around 0.4, so a chance coincidence cannot be robustly ruled out. 

\revise{CX7-BY1 and CX7-BY2 might point to an AB nature; however, the relatively high X-ray/optical ratio strongly argues against this interpretation (Figure \ref{fig:x-ray-to-optical-ratio}). Considering CX7's large offset distance from the cluster centre, it is also possible that CX7 is a background AGN; this is reasonable considering that the probability of finding at least one AGN within CX7's radial offset is around $0.97$ (Figure \ref{fig:n_agn_vs_r}; Sec \ref{sec:agn-number_estimate}).} 

Our identification will be further complemented by future observations. Typically for CX5 and CX7, where red outliers' $\nc$ values are not sufficient in ruling out chance coincidences, future spectroscopic follow-ups or imaging observations with H$\alpha$ photometry will be useful in reducing the degeneracy.

\subsubsection{PSR J1737--0314A (M14 A)}
The epoch of M14 A's timing position is MJD 58900 \citep{Pan21}, which is 4.14\,years after the \gaia\ epoch. We therefore backtrack its position to the \gaia\ epoch using the cluster proper motion from \citet{Vasiliev21}. This gives $0\farcs015$ and $0\farcs02$ shifts (\gaia - timing) in RA and DEC, respectively. The corrected position is in the vicinity of the main sequence source R0102981 in the \citetalias{DAntona22} catalogue (Figure \ref{fig:hst-finders}). \revise{M14 A was found to be a black widow pulsar with a minimum companion mass of $0.016\,\Msun$ in a close orbit \citep[$\approx 5.5\,\mathrm{hour}$;][]{Pan21}. In such a close orbit, the low-mass companion is tidally locked with the pulsar, and the side facing the pulsar is strongly irradiated and heated by the pulsar wind \citep[e.g.,][]{Romani16}. Irradiation can significantly increase the brightness of the low-mass companion, possibly reaching that of R0102981 \citep[e.g.,][]{Draghis19, Koljonen23}; however, a major counterargument arises from the lack of variability observed in the \hst\ bands (Figure \ref{fig:rms-mag-plot}), while variations are expected in black widow systems as the heated side and the ``night" side of the companion alternate within the line of sight.} 

% Typical black widow pulsars have companion masses around $0.02$ \citep{Fruchter88, Stovall14}, which are below the detection limit of the \hst\ observation.

The radio timing position is also marginally consistent with the \vla\ source VLA45, which is only detected in the $\nulo$ sub-band, so $\alpha$ is only constrained by an upper limit ($\alpha < 0$); however, given the small timing and \vla\ positional uncertainties, we argue that VLA45 is very unlikely a chance coincidence with M14 A.

\citet{ZhaoJ22} extracted the X-ray spectrum of M14 A, from a circular region centred at its timing position, and found its spectrum is well fit by a \bbody\ model with an X-ray luminosity (0.5--10 keV) of $\approx4\times10^{31}\,\ergps$. M14 A is therefore the most X-ray-luminous black widow pulsar in GCs found to date, with unusually substantial thermal emission. However, it is noteworthy that there were only five photons extracted in the spectrum of M14 A, leading to large uncertainty in the spectral fitting. Hence, deeper X-ray observations of M14 are required to better resolve its spectrum and constrain its X-ray properties.

\subsection{Other MSPs in M14}
\label{sec:other-msps-in-m14}
Apart from M14 A, another four MSPs have been detected in M14, although the others do not yet have available timing positions. M14 B and C are found in binary systems and both have an orbital period of $\approx 8.5\,\mathrm{day}$ \citep{Pan21}, indicating that pulsar wind is unlikely to as strongly interact with the companion as in spider pulsars; the latter typically have orbital periods less than 1 day but could reach up to 1.97 days \citep[NGC 6397 B;][]{Zhang22}. Such canonical MSP binaries typically have X-ray luminosities less than $10^{31}$ $\ergps$ \citep{Bogdanov06,ZhaoJ22}, while our limiting X-ray luminosity for 
M14 is about $6.5\times10^{31}\,\ergps$ \citepalias{Bahramian20}. Therefore, the X-ray counterparts to M14 B and C are not expected to be detected in this \chandra\ observation. 
On the other hand, M14 D and E are found to be eclipsing redbacks, with minimum companion masses of $0.13\,\msun$ and $0.17\,\msun$, and orbital periods of 0.74 days and 0.85 days, respectively \citep{Pan21}.
A recent work by \citet{ZhaoJ23} reveals a positive correlation between X-ray luminosities and minimum companion masses for spider pulsars. According to their results, the X-ray luminosities of M14 D and E are predicted in ranges of [0.2, 1.2]$\times10^{32}\,\ergps$ and [0.2, 1.7]$\times10^{32}\,\ergps$, respectively, in 0.5--10 keV. 

\revise{Can some of our X-ray sources be MSPs? CX2 has an X-ray luminosity (0.5--10 keV) of \revise{$3.2^{+1.4}_{-0.9}\times 10^{32}\,\ergps$ (based on the best-fit model in Table \ref{tab:spectral_fit_results})}, which is higher than either of the predicted luminosities of M14 D and E (it is consistent with the prediction for M14 E within 2-$\sigma$ though). 
%If CX2 is indeed a MSP, it will be the second brightest MSP and the brightest redback pulsar in X-rays in GCs found to date. 
%Plausibly, it might be a transitional MSP observed in the pulsar or low-luminosity `passive' state \citep[as observed for M28 I; see e.g.][]{Papitto13,Linares14}. 
CX3 might be another redback, considering its X-ray luminosity and X-ray colour. However, as discussed above, its location outside the core increases the possibility of an AGN scenario. CX4's exotic SSG counterpart could be reminiscent of redback MSPs in other clusters \citep[e.g., NGC 6397 A and B;][]{Zhao20a}, despite a relatively low X-ray/optical ratio. CX5's relatively soft X-ray spectrum argues against spider pulsars that typically have hard  non-thermal spectra, %l components, 
but it could be %interpreted by
similar to 
pulsars in wider binaries such as M14 B or C. Finally, the very faint CX6 and CX7 do not provide very strong constraints on their X-ray spectra, but an AGN interpretation is plausible considering their relatively large offsets from the cluster centre. In all, more conclusive insights will be drawn from future radio timing solutions of these MSPs and X-ray observations of M14.}

\subsection{AGN number estimate}
\label{sec:agn-number_estimate}
\begin{figure}
    \centering
    \includegraphics[width=\columnwidth]{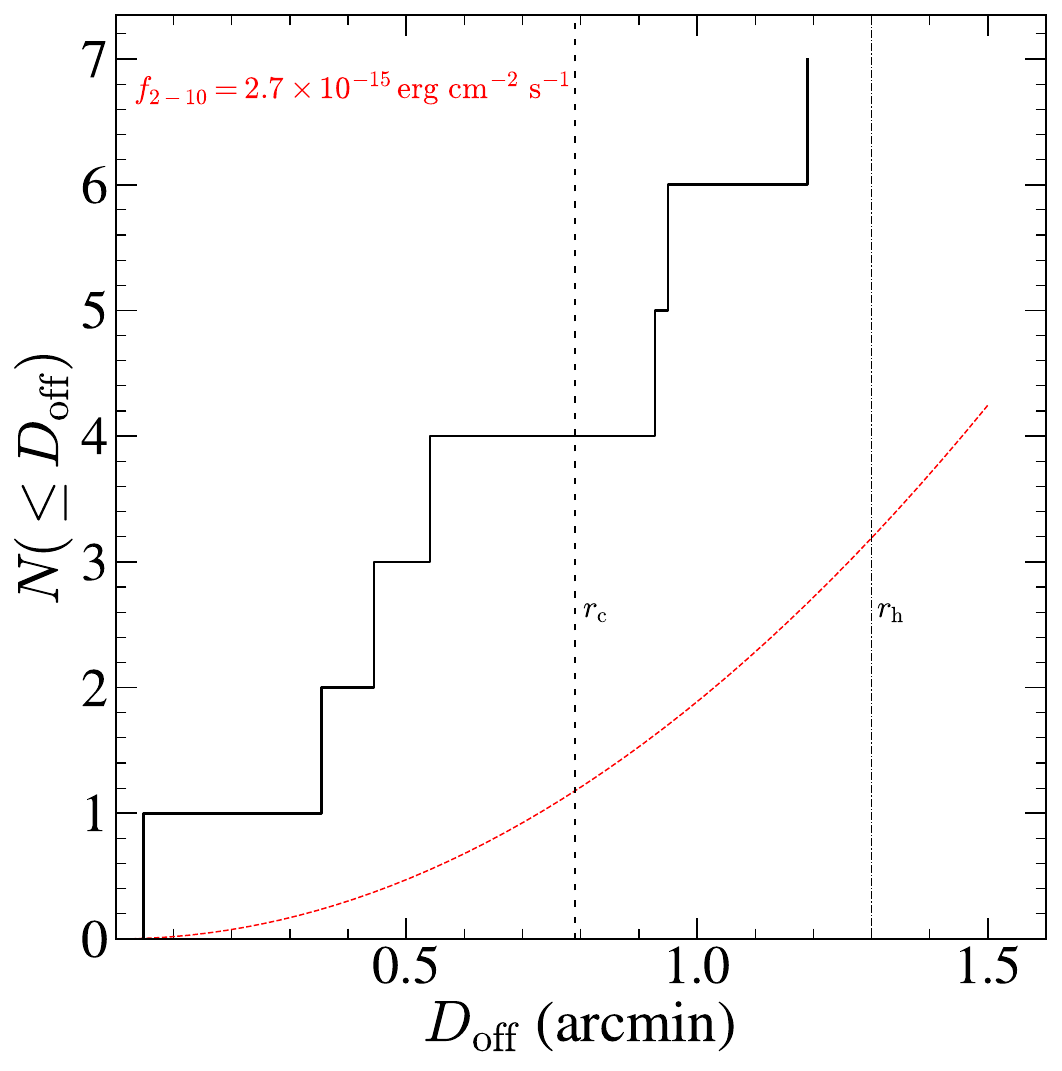}
    \caption{Empirical cumulative distribution function (ECDF) of X-ray source offsets (solid black line) compared to expected number of AGNs (dashed red line) as a function of radial offset from the cluster centre. }
    \label{fig:n_agn_vs_r}
\end{figure}

Sources further away from the cluster centre are more likely to be background AGNs simply because the cumulative number of AGNs increases with larger sky area, 
while the number density of true cluster sources declines. 
We estimate the number density of AGNs using the empirical model in \citet{Mateos08}; specifically, we convert the model $\fhard$ of CX7 to $2-10\,\kev$ to match the band in \citet{Mateos08} and use this as the flux limit. The number of AGN is then estimated as a function of radial offset from cluster centre and presented in Figure \ref{fig:n_agn_vs_r}; as a comparison, we also plot the empirical cumulative distribution function (ECDF) of source offsets. The source excess at smaller offsets is likely to be more significant. %but this could only be more robustly confirmed with higher counting statistics. 
As a measure of probability, we also calculate the Poisson probability of finding at least $1$ AGN within the given radial offset ($\probagn$), given by $1 - \exp(-N_\mathrm{AGN})$ (Table \ref{tab:hst_counterparts}). Overall, we predict $\approx 3$ AGNs within $\rh$ that have $2-10\,\kev$ flux above $2.7\times 10^{-15}\,\ergscm$. %so we expect at least some of our sources to be members. 
Most likely, 3 of the 4 sources within $r_c$ are cluster members, and 1 of the 3 sources between $r_c$ and $r_h$.

\section{Conclusions}
\label{sec:conclusions}
Our study of the 12.09\, ks \chandra\ observation of the massive GC M14 leads to the detection of 7 X-ray sources within its half-light radius at a $0.5-7\,\kev$ depth of $2.5\times 10^{31}\,\ergps$. Cross-matching with an \hst\ photometry catalogue and the MAVERIC radio source catalogue reveals \hst\ counterparts to 6 and \vla\ counterparts to 2 of the X-ray sources. We find that both CX1 and a UV-bright variable source are consistent with the nominal position of classical nova Oph 1938, making it the second classic nova recovered in a Galactic GC after Nova T Sco in M80. CX2 is consistent with the steep radio source VLA8 whose position matches a \hst\ source with clear UV and blue excess. \referee{We thus consider CX2 a possible transitional millisecond pulsar based on its radio steepness, so the X-ray and UV/optical/IR observations were made during its accretion-powered LMXB phase, while the steep radio source was observed during a pulsar phase.} Another X-ray source that matches a radio source position is CX3, but the radio source (VLA50) has an unconstrained spectral index, so its nature is less certain. The other \hst\ counterparts with faint X-ray sources point to identifications as likely ABs or AGNs, but further membership information is required to exclude interlopers. There is also a radio source consistent with the timing position of the recently discovered MSP M14 A, highlighting the MAVERIC survey's potential to uncover new pulsars. \revise{Additionally, other faint X-ray sources could be associated with M14 D and E --- two recently discovered redback MSPs, but further timing positions are needed before more conclusive identifications are made.}

\section*{Acknowledgements}

% This research has made use of data obtained from the Chandra Data Archive and the Chandra Source Catalog, and software provided by the Chandra X-ray Center (CXC) in the application packages CIAO and Sherpa.
We thank the anonymous reviewer for their helpful comments. YZ acknowledge the Science and Technology Facilities Council grant ST/V001000/1 for support. JZ is supported by China Scholarship Council (CSC), File No. 202108180023. The scientific results reported in this article are based on data obtained from the Chandra Data Archive. Data processing and analysis have made use of the software provided by the Chandra X-ray Center (CXC) in the application packages {\sc ciao} and {\sc sherpa}; retrieving and calibrating \gaia\ BP/RP spectrum has made use of the Python package {\sc GaiaXPy}, developed and maintained by members of the Gaia Data Processing and Analysis Consortium (DPAC), and in particular, Coordination Unit 5 (CU5), and the Data Processing Centre located at the Institute of Astronomy, Cambridge, UK (DPCI). Finally, this research has also made use of the VizieR catalogue access tool, CDS, Strasbourg, France. 

In addition to the software mentioned in the texts, analysis and visualisation in this work have also made use of the following packages (in alphabetic order): {\sc astropy}:\footnote{http://www.astropy.org} a community-developed core Python package and an ecosystem of tools and resources for astronomy \citep{Astropy22}, {\sc matplotlib} \citep{Matplotlib07}, {\sc numpy} \citep{Numpy20}, {\sc pandas} \citep{Pandas23}, {\sc photutils}, an {\sc astropy} package for
detection and photometry of astronomical sources \citep{Bradley23}, and {\sc scipy} \citep{Scipy20}.

%%%%%%%%%%%%%%%%%%%%%%%%%%%%%%%%%%%%%%%%%%%%%%%%%%
\section*{Data Availability}
The data used in this work are publicly available and can be queried using web-based portals. The \chandra\ data (observation ID: 8947) can be downloaded from the Chandra Data Archive (Sec \ref{sec:x-ray-observations}). \hst\ imaging data associated with the proposal ID 16283 can be searched and downloaded from the Mikulski Archive for Space Telescopes (Sec \ref{sec:hst-observations}). The MAVERIC catalogue has been published and can be queried using the VizieR catalogue access tool, and the associated \vla\ data can be downloaded from the NRAO data archive\footnote{\url{https://science.nrao.edu/facilities/vla/archive/index}}.

% The inclusion of a Data Availability Statement is a requirement for articles published in MNRAS. Data Availability Statements provide a standardised format for readers to understand the availability of data underlying the research results described in the article. The statement may refer to original data generated in the course of the study or to third-party data analysed in the article. The statement should describe and provide means of access, where possible, by linking to the data or providing the required accession numbers for the relevant databases or DOIs.



\bibliographystyle{mnras}
\bibliography{ref} % if your bibtex file is called example.bib

%\begin{thebibliography}{99}
%\end{thebibliography}

%%%%%%%%%%%%%%%%%%%%%%%%%%%%%%%%%%%%%%%%%%%%%%%%%%

%%%%%%%%%%%%%%%%% APPENDICES %%%%%%%%%%%%%%%%%%%%%

\appendix

\section{Off-axis sources used for boresight correction}

\begin{table*}
\centering
\caption{Off-axis sources used for absolute astrometry.}
    \begin{tabular}{llccccccc}
        \toprule
        ID & source\_id & RA (\chandra) & DEC (\chandra) & RA (\gaia) & DEC (\gaia) & $\varpi^a$ & \gaia\ G & BP$-$RP \\
        & & (hh:mm:ss.ss) & $^\circ$:$\arcmin$:$\arcsec$ & (hh:mm:ss.ss) & $^\circ$:$\arcmin$:$\arcsec$ & ($\mas$) & & \\
        \midrule
		A (YSO) & 4368928767444987648 & 17:37:48.42 & $-$03:15:47.77 & 17:37:48.41 & $-$03:15:47.95 & $2.31\pm 0.07$ & $13.4$ & $1.5$ \\
		B$^b$ & 4368928767444986880 & 17:37:48.22 & $-$03:15:52.96 & 17:37:48.24 & $-$03:15:53.15 & $2.13\pm 0.02$ & $12.5$ & $0.7$ \\
        \bottomrule
        \multicolumn{9}{l}{$^a$\gaia\ parallaxes corrected for zero-point offsets \citep{Lindegren21}.}\\
        \multicolumn{9}{l}{$^b$\gaia\ magnitude and colour have been corrected for extinction and reddening.}
    \end{tabular}
    \label{tab:offaxis_sources}
\end{table*}

\begin{figure}
    \centering
    \includegraphics[width=\columnwidth]{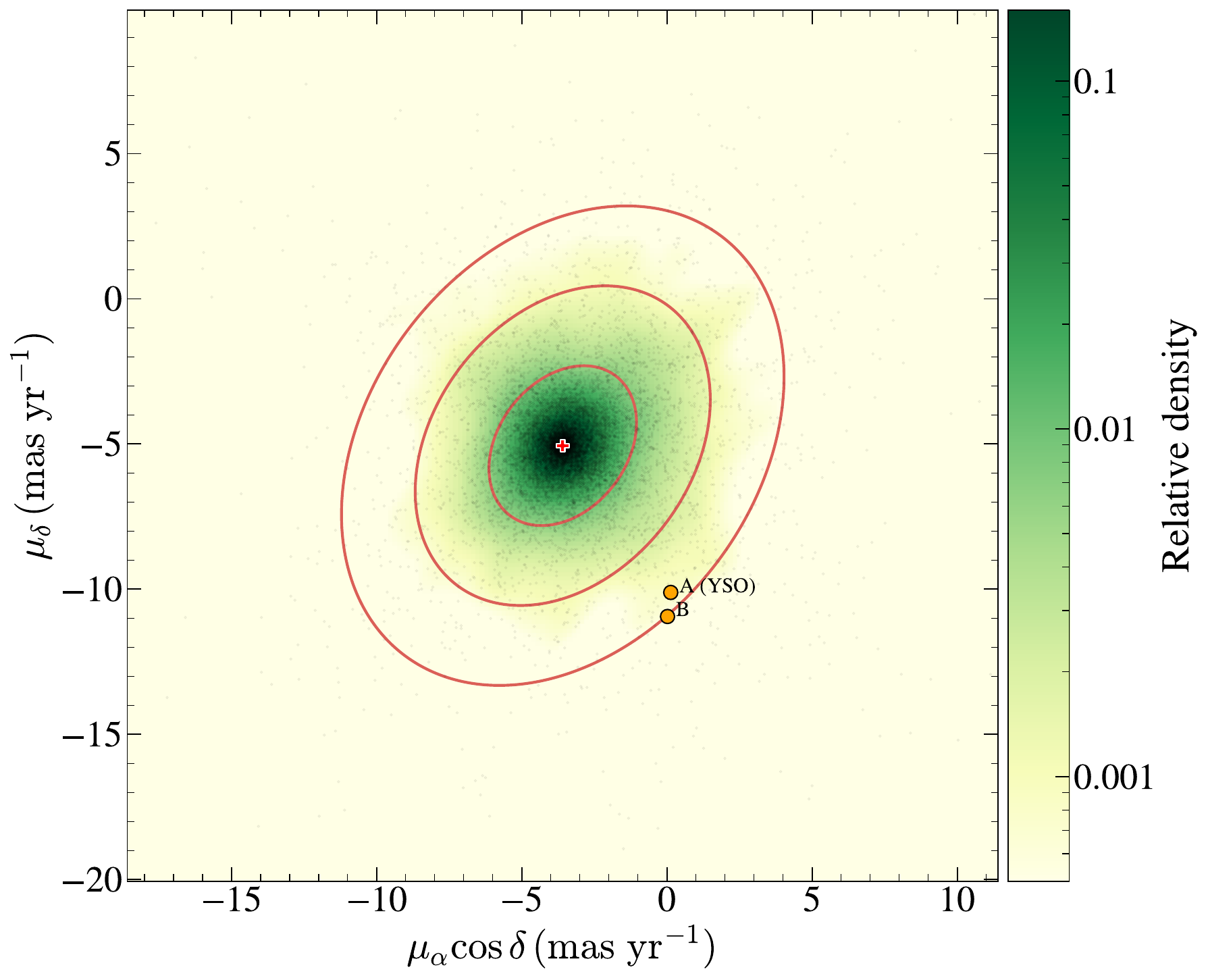}
    \caption{A VPD showing the proper motion components in RA and DEC for \gaia\ sources within $4\rh=5\farcm2$ of M14 centre. The colour scale represents a density estimate using a Gaussian kernel. The red contours indicate (from the innermost to the outermost) 1, 2, and 3\,$\sigma$ confidence ellipses centered on the nominal cluster proper motion (marked by a red cross) from \citet{Vasiliev21}. The ellipses are calculated assuming the scatter follows a 2D Gaussian distribution. The two filled orange circles represent the \gaia\ counterparts to the two off-centre X-ray sources used for boresight correction (Sec \ref{sec:x-ray-astrometry}) --- both are at $\approx 3\sigma$ away from the cluster proper motion.}
    \label{fig:gaia_vpd}
\end{figure}

\begin{figure*}
    \centering
    \includegraphics[width=0.8\textwidth]{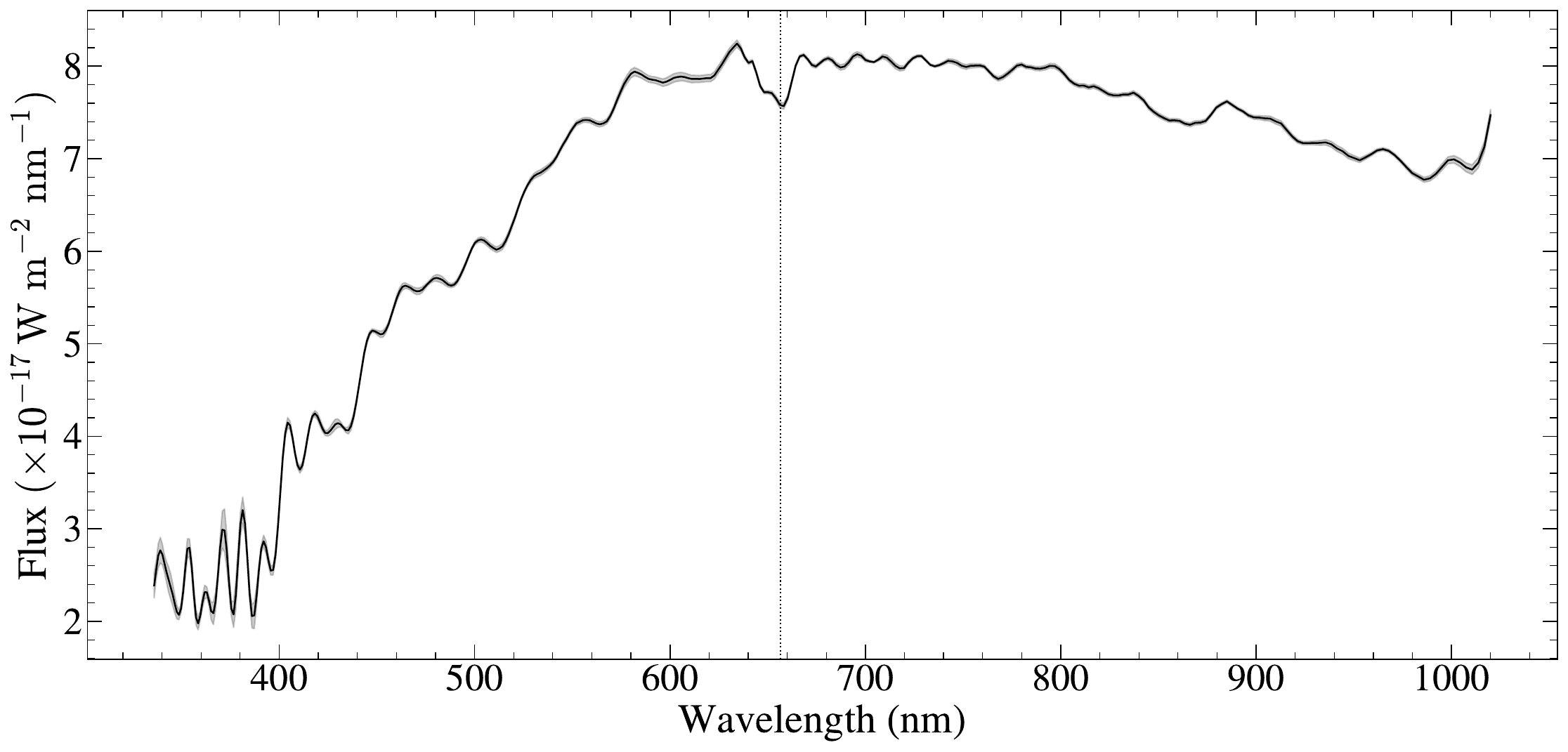}
    \caption{Gaia BP/RP spectrum of 436892876744498688 (the \gaia\ counterpart to source B in Figure \ref{fig:chandra_offaxis_image}). The vertical line marks the vacuum wavelength of the H$\alpha$ line, around which the spectrum exhibits a likely broadened absorption feature.}
    \label{fig:bp-rp-spectrum}
\end{figure*}

%%%%%%%%%%%%%%%%%%%%%%%%%%%%%%%%%%%%%%%%%%%%%%%%%%

% Don't change these lines
\bsp	% typesetting comment
\label{lastpage}
\end{document}